\documentclass[10pt,a4paper]{article}
\pdfoutput=1
\usepackage[utf8]{inputenc}
\usepackage{authblk}
\usepackage{amsmath,bm}
\usepackage{amsfonts}
\usepackage{amssymb}
\usepackage{graphicx}
\usepackage{subcaption}
\usepackage{geometry}

\usepackage{hyperref}
\hypersetup{colorlinks=true,allcolors=blue}
\usepackage{hypcap}

\newcommand{\andothers}{\emph{et al.} }
\newcommand*\paris{\textsc{PARIS} }

\newcommand*\code[1]{\textsc{#1} }
\newcommand{\mdotS}{\dot{m}^{\prime\prime} }

\newcommand*\rfrac[2]{{}^{#1}\!/_{#2}}

\newcommand*\nb{\bm{\nabla}^h}

\title{A Geometric VOF Method for Interface Resolved Phase Change and Conservative Thermal Energy Advection}

\author[a,b,c]{L.C. Malan} 
\author[c]{A.G. Malan}
\author[a,b]{S. Zaleski}
\author[d]{P.G. Rousseau}

\affil[a]{Sorbonne Universit\'{e}s, UPMC Univ Paris 06, F-75005, Paris, France}
\affil[b]{CNRS, UMR 7190, Institut Jean le Rond d'Alembert, F-75005, Paris, France}
\affil[c]{Industrial CFD Research Group, Dept. Mechanical Engineering, University of Cape Town, Private Bag X3, Rondebosch 7701, South Africa}
\affil[d]{ATPRoM Research Group, Dept. Mechanical Engineering, University of Cape Town, Private Bag X3, Rondebosch 7701, South Africa}

\begin{document}

\maketitle

\begin{abstract}

We present a novel numerical method to solve the incompressible Navier-Stokes equations for two-phase flows with phase change, using a one-fluid approach. Separate phases are tracked using a geometric Volume-Of-Fluid (VOF) method with piecewise linear interface construction (PLIC). Thermal energy advection is treated in conservative form and the geometric calculation of VOF fluxes at computational cell boundaries is used consistently to calculate the fluxes of heat capacity. The phase boundary is treated as sharp (infinitely thin), which leads to a discontinuity in the velocity field across the interface in the presence of phase change. The numerical difficulty of this jump is accommodated with the introduction of a novel two-step VOF advection scheme. The method has been implemented in the open source code \paris and is validated using well-known test cases. These include an evaporating circular droplet in microgravity (2D), the Stefan problem and a 3D bubble in superheated liquid. The accuracy shown in the results were encouraging. The 2D evaporating droplet showed excellent prediction of the droplet volume evolution as well as preservation of its circular shape. A relative error of less than 1\% was achieved for the Stefan problem case, using water properties at atmospheric conditions. For the final radius of the bubble in superheated liquid at a Jacob number of 0.5, a relative error of less than 6\% was obtained on the coarsest grid, with less than 1\% on the finest.

\end{abstract}

\section{Introduction}

A popular approach used in the wider problem space of the numerical solution of multiphase-flows is the so-called one-fluid formulation, in which the governing equations (incompressible Navier-Stokes equations for two-phase flow in our case) are applied to both phases \cite{Tryggvason05}. Some phase tracking technique is then employed to calculate the phase-dependent fluid properties. For an overview of different interface tracking techniques and associated methods to account for phase change, the reader can refer to the work of Kharangate \& Mudawar \cite{Chirag17} and Tryggvason \andothers \cite{Tryggvason05}. 

In this work, the focus will be on the Volume-of-Fluid (VOF) method \cite{Hirt81}. Even though VOF methods are relatively common for incompressible, isothermal two-phase flows, very few include temperature driven phase change. Some of the pertinent VOF works dealing with interface resolved phase change problems will be mentioned here as a reference. Welch and Wilson \cite{Welch00} developed a VOF method for phase change on staggered (MAC) grids. The interface is reconstructed using piecewise linear segments, based on the method of Youngs \cite{Youngs82}. A time-split geometric advection technique, also by Youngs, is used to advect the interface. The energy conservation equation is treated non-conservatively. Welch and Rachidi \cite{Welch02} expanded this method to study film boiling with conjugate heat transfer. Agarwal \andothers \cite{Agarwal04} used the Welch method \cite{Welch00} with the addition of temperature dependence on fluid properties. Schlottke and Weigand used a geometric based method to study evaporating droplets \cite{Schlottke08}. They developed an intricate, iterative method to deal with the calculation of a volume source due to mass transfer, which they apply only in interface cells (sharply). However, they neglect the velocity disparity between the interface and the liquid phase. 

Kunkelmann \cite{Kunkelmann11} used an algebraic VOF advection technique in the OpenFOAM framework and implemented a microlayer and contact angle model to simulate nucleate boiling on unstructured meshes. The mass transfer between phases is accounted for using a smearing technique: source terms are redistributed in some finite region around the interface, which smoothes out the sharp discontinuity in velocity at the interface. Guedon \cite{GuedonThesis} also used the OpenFOAM framework for arbitrary meshes, but attempted to apply a sharp interface approach instead of the smearing technique used by Kunkelmann. A ghost-fluid approach was used at the interface to apply the jump conditions. The author, however, reported difficulties from spurious currents due to unbalanced surface tension models. Another difference compared to the method of Kunkelmann, is the non-conservative treatment of energy conservation.

In summary, there remains several non-trivial challenges to the numerical treatment of the discontinuities at the interface in VOF phase change modelling schemes. More specifically, the authors are unaware of a method that combines conservative thermal energy advection with a sharp interface treatment of the velocity discontinuities that arise from phase change. This is particularly pressing due to the large heat capacity ratio (per unit volume) between liquid and gas phases of many substances of practical interest.

In this work we aim to address the above challenges. A novel method is developed which simulates the incompressible flow of pure substances undergoing phase change. A VOF method with piecewise linear interface calculation (PLIC) is employed. The interface is treated as sharp (infinitely thin), separating immiscible fluids. A novel geometric VOF advection method is developed to deal with the discontinuity in the velocity that arises from phase change. It is applied consistently to the calculation of the thermal energy advection term, which is treated in conservative form. The method is designed such that it can be easily incorporated into an existing VOF framework using geometric advection algorithms, such as that of Weymouth and Yue \cite{Weymouth10}.

\section{Mathematical Formulation} \label{sec:maths}

\begin{figure}[ht]
	\centering
	\def\svgwidth{0.75\textwidth}
	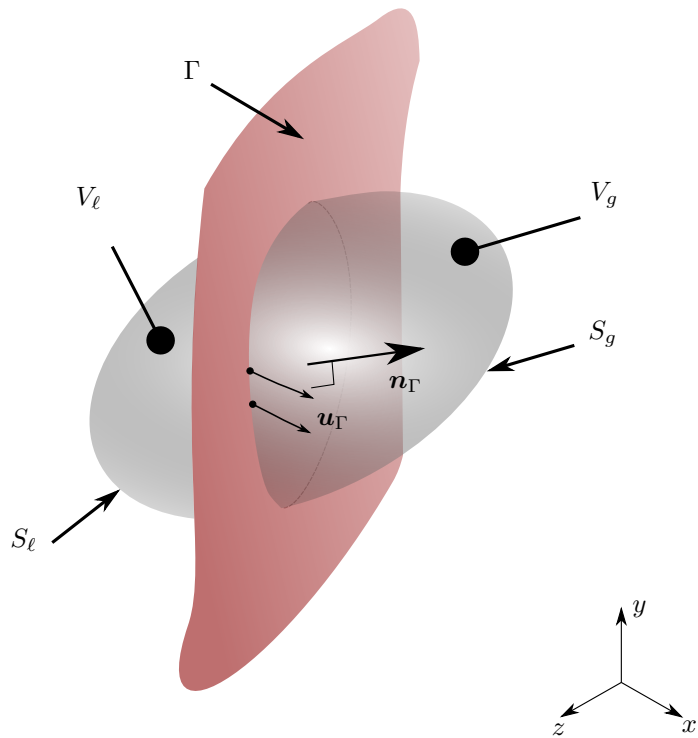
	\caption{The control volume under consideration, $V$, contains a sharp interface $\Gamma$ (red shading) and is bounded in space by fixed surfaces $S_\ell$ and $S_g$. The interface has velocity $\bm{u}_\Gamma$, which may differ from the velocity of the adjacent fluid.}
	\label{fig:cv_V_P2}
\end{figure}

This work entails modelling the incompressible flow of pure substances undergoing phase change and in this section we describe the integral form of the governing equations as applied in a generic control volume, $V$. For a detailed derivation, refer to \cite{PhDLMalan}. Fig.~\ref{fig:cv_V_P2} shows volume $V$, which is fixed in space and bounded by surface $S$ of arbitrary shape with outward pointing normal $\bm{n}$. Surface $\Gamma$ is assumed to have zero thickness, intersects $V$ and represents the phase boundary (interface) between two pure phases. Note that we will not consider molecular diffusion between phases in this work and will name the two phases liquid ($\ell$) and gas ($g$) throughout. The surface $\Gamma$ has a normal $\bm{n}_\Gamma$ which points from the liquid into the gas phase. We assume constant material properties (density, viscosity and heat conductivity) in the respective liquid and gas phases.

For phase change at the interface, $\Gamma$, we define $\mdotS$ as the rate at which liquid is converted to gas (or vapour). It has units of mass per unit area per unit time, so that 

\begin{equation}
\mdotS = \rho_\ell \left(\bm{u}_\ell - \bm{u}_{\Gamma} \right) \cdot \bm{n}_{\Gamma} = \rho_g \left(\bm{u}_g - \bm{u}_{\Gamma} \right) \cdot \bm{n}_{\Gamma} \,.
\label{eq:mdot}
\end{equation}
Here $\bm{u_\ell}$ and $\bm{u_g}$ are the respective liquid and gas phase velocities adjacent to the interface. The velocity of $\Gamma$ is $\bm{u}_\Gamma$.

We use the Heaviside function, $H$, as an indicator for the liquid phase, so that:

\begin{equation}
H(\bm{x},t) = 
\begin{cases}
1, & \text{if } \, \bm{x} \, \text{ is inside } \, V_\ell \\
0, & \text{if } \, \bm{x} \, \text{ is inside } \, V_g \,.
\end{cases}
\label{eq:Heaviside}
\end{equation}
The definition of $H$ can be constructed by using the integral of consecutive products of one dimensional Dirac delta functions, which is extended to three dimensions for $V$ from the derivation in \cite[~p.279]{Book11}.

The rate of change of $H$ in $V$ is given by

\begin{equation} \label{eq:cons_Heavy}
\int_V \dfrac{\partial H}{\partial t} \, dV + \oint_S H \bm{u} \cdot \bm{n} \, dA + \int_\Gamma \left(\bm{u}_\ell - \bm{u}_{\Gamma} \right) \cdot \bm{n}_{\Gamma}dA = 0\,,
\end{equation}
where the last term on the left hand side accounts for phase change at the interface along with the previously defined convention that $H=1$ in the liquid phase.

The VOF represents the volumetric fraction of liquid in $V$. We will denote it $c$, from the popular usage of \emph{colour} function

\begin{equation} \label{eq:VOF_def}
c = \dfrac{1}{V} \int_{V} H \left(\bm{x},t \right) \, dV \,.
\end{equation}
From the equation above, it can be deduced that $0 \leq c \leq 1$. For $V$ completely filled with liquid $c=1$ and $c=0$ when only gas is present. Note that $c$ is a scalar function of which the definition is always relative to the volume in which it is defined. Using \eqref{eq:mdot}, \eqref{eq:cons_Heavy} and \eqref{eq:VOF_def}, we can write for $c$ that

\begin{equation} \label{eq:vof_advect_mdot}
\dfrac{\partial c}{\partial t} + \nabla \cdot \left( \bm{u} c \right) + \dfrac{\mdotS}{\rho_{\ell}} \delta_\Gamma = 0 \,,
\end{equation}
where $\delta_\Gamma$ is a Dirac delta function defined on $\Gamma$. Equation \eqref{eq:vof_advect_mdot} is intended in a weak sense, since $c$ is not a continuous function and is solved by integrating over a volume. Let $\overline{\phi}$ be some volume averaged material property. We then have that

\begin{equation} \label{eq:c_vol_avg}
\overline{\phi} = \phi_\ell c + (1-c)\phi_g
\end{equation}
with the subscripts indicating the phase. We can now write a single set of the governing equations, also referred to as the ``one-fluid'' approach. The fluid density ($\rho$), kinematic viscosity ($\mu$) and heat capacity ($C_p=\rho c_p$) is calculated using \eqref{eq:c_vol_avg}, with the product of density and specific heat capacity creating a heat capacity per unit volume.

From conservation of mass in $V$ we have that 
  
\begin{equation} \label{eq:masscons_int}
\int_{V} \bm{\nabla} \cdot \bm{u} \, d V = \left( \frac{1}{\rho_g} - \frac{1}{\rho_\ell} \right) \int_{V} \mdotS \delta_\Gamma\, dV \,.
\end{equation}
Here $\bm{u}$ is the fluid velocity.

The conservation of momentum is written

\begin{equation} \label{eq:momcons_int}
\int_{V} \dfrac{\partial \left( \rho \bm{u} \right)}{\partial t} \, dV + \oint_{S} \rho \bm{u} \left( \bm{u} \cdot \bm{n} \right) \, dA = \oint_{S} \bm{\tau} \cdot \bm{n} \ dA + \int_{V} \rho \bm{g} \ d V + \int_{V} \sigma \kappa \delta_\Gamma \bm{n}_\Gamma \, dV \,,
\end{equation}
with $\bm{\tau}$ the stress tensor, which for our problem is

\begin{equation}
\bm{\tau} = -p\bm{\mathrm{I}} + \mu \left( \nabla \bm{u} + \left(\nabla \bm{u}\right)^{T}\right) \,.
\end{equation}
Here $p$ is the static pressure, $\bm{\mathrm{I}}$ the unit tensor and $\mu$ the kinematic viscosity. The last integral in \eqref{eq:momcons_int} represents the surface tension force on $\Gamma$ with $\sigma$ the surface tension coefficient and $\kappa$ the curvature of $\Gamma$.

Conservation of thermal energy is given by
\begin{equation} \label{eq:etcons_int}
\int_{V} \frac{\partial \left( C_p T \right) }{\partial t} \, dV + \oint_S C_p T \bm{u} \cdot \bm{n} \, dA = \oint_{S} k \bm{\nabla} T \cdot \bm{n} \, dA - \int_V \dot{q}_\Gamma \delta_\Gamma \, dV \,,
\end{equation}
with the thermal energy flux term in the second integral calculated conservatively in this work, contrary to many approaches where that integral is written as $\overline{C_p} \oint_S T \bm{u} \cdot \bm{n} \, dA$. $T$ is the temperature and $k$ the thermal conductivity coefficient. The last term in \eqref{eq:etcons_int} is a thermal energy source located on the interface due to phase change latent heat transfer

\begin{equation} \label{eq:et_jump}
\dot{q}_\Gamma = \mdotS h_{fg} = k_g \nabla_{n_\Gamma} T_g - k_\ell \nabla_{n_\Gamma} T_\ell \,,
\end{equation}
with $h_{fg}$ the latent heat of evaporation and $\nabla_{n_\Gamma} T$ the normal temperature gradient on $\Gamma$ in the respective phase. 

The above equations are discretized on a regular Cartesian grid prior to an iterative numerical solution process. This is described next.

\section{Numerical Method}

The system of equations described in the previous section is solved numerically on a marker-and-cell (MAC) type computational mesh (also known as a staggered grid), created by Harlow and Welch \cite{Harlow65}. Discrete velocity components are located on cell faces, while other scalar variables -- like pressure, VOF and temperature -- are located at cell centres. The time integration procedure will first be presented to provide an overview of the scheme, after which novel numerical techniques will be detailed. These include a PLIC based mass transfer rate calculation, a novel method of dealing with interface velocity jump conditions while maintaining a sharp interface definition and the discretisation of the thermal energy advection in conservative form using consistent VOF flux calculations for thermal heat capacity. 

\subsection{Solution overview} \label{sec:time_integration_full}
 
We next detail the time integration procedure, assuming that all variables are known at some time step $n$ for all computational cells and initial values provided when $n=0$. Time integration is done with an explicit, first order (forward Euler) scheme, using a split pressure projection method \cite{Chorin69}. 

\begin{enumerate}

\item The forward Euler time integration of the thermal energy conservation equation is first applied to each cell and reads

\begin{equation} \label{eq:et_disc}
C_pT\vert^{n+1}=C_pT\vert^n + \Delta t \left[ \nb \cdot k \nb T\vert^{n+1} - \nb \cdot \left( C_pT\bm{u} \right)\vert^n - \dot{q}\delta_\Gamma\vert^n \right] \,,
\end{equation}
where $\Delta t$ is the discrete time step. The $h$ superscript for the gradient operator indicates that it is defined in a discrete sense. The diffusion of thermal energy is calculated implicitly and will be detailed in Section \ref{sec:et_diff}. Calculation of the thermal energy advection term will be discussed in more detail in Section \ref{sec:et_adv}. The last term deals with latent energy transfer due to phase change at the interface and is calculated from \eqref{eq:et_jump}, with the mass transfer rate at the interface, $\mdotS$, calculated from the temperature field, $T^n$  

\begin{equation}
\mdotS = \dfrac{1}{h_{fg}} \left( k_g\nabla_{n_\Gamma}^h T\vert^n_g - k_\ell \nabla_{n_\Gamma}^h T\vert^n_\ell \right) \,.
\label{eq:discrete_mdot}
\end{equation}
This is done for all cells where $\epsilon<c<1-\epsilon$, with $\epsilon=10^{-12}$ in this work. The method used to calculate this term is described in Section \ref{sec:heat_flux}.

\item The VOF equation is integrated in time as 

\begin{equation} \label{eq:VOF_num}
\dfrac{c^{n+1} - c^n}{\Delta t} + \bm{\nabla}^h \cdot \left( \bm{u} c \right)\vert^n + \dfrac{\mdotS}{\rho_{\ell}} \delta_\Gamma\vert^n = 0 \,.
\end{equation}
This equation is solved numerically using a geometric (PLIC) method. The procedure is detailed in Section \ref{sec:VOF_adv}. All fluid properties that depend on $c$ can now be calculated at time $n+1$ ($\rho^{n+1}$, $\mu^{n+1}$, $C_p^{n+1}$, $\kappa^{n+1}$, $\delta_\Gamma^{n+1}$, $\bm{n}^{n+1}_\Gamma$).

\item The predicted velocity $\bm{u}^{\ast}$ in the momentum equation is calculated from
\begin{equation}
\dfrac{\bm{u}^{\ast}-\bm{u}^{n}}{\Delta t}= - \bm{u}^{n} \cdot \bm{\nabla}^{h} \bm{u}^n + \dfrac{1}{\rho^{n+1}} \left( \bm{\nabla}^{h} \cdot 2 \mu \bm{\mathrm{S}} + \sigma \kappa \delta_\Gamma \bm{n}_\Gamma\vert^{n+1} \right) + \bm{g} \, . 
\label{eq:u_temp_full_num}
\end{equation}
Here $\bm{\mathrm{S}}$ is the fluid strain rate tensor, given by $\bm{\mathrm{S}}=\rfrac{1}{2} \left( \nabla \bm{u} + \left(\nabla \bm{u}\right)^{T}\right)$. Note that the viscous term can be calculated explicitly ($\bm{\nabla}^{h} \cdot 2 \mu \bm{\mathrm{S}}\vert^{n}$) or implicitly ($\bm{\nabla}^{h} \cdot 2 \mu \bm{\mathrm{S}}\vert^{n+1}$). For more detail on the discretization of this equation, refer to \cite{PARIS19}.

\item A pressure Poisson equation is then solved for the pressure at $n+1$

\begin{equation}
\bm{\nabla}^{h} \cdot \left[\frac{\Delta t}{\rho^{n+1}} \bm{\nabla}^{h} p^{n+1} \right] = \bm{\nabla}^{h} \cdot \bm{u}^{*} - \mdotS \left( \dfrac{1}{\rho_g} - \dfrac{1}{\rho_\ell}\right) \delta_\Gamma^n
\label{eq:press_full_num}
\end{equation}

\item The one fluid velocity $\bm{u}^{n+1}$ is obtained with a correction step using $p^{n+1}$

\begin{equation}
\dfrac{\bm{u}^{n+1}-\bm{u}^{*}}{\Delta t} = -\dfrac{1}{\rho^{n+1}}\bm{\nabla}^{h} p^{n+1}
\label{eq:correct_full_num}
\end{equation}

\end{enumerate}

All variables have now been solved at time step $n+1$ and the cycle can repeat for the next time step. The numerical method presented here has been implemented in \paris with full parallel processing capability for three dimensional flow.

\subsection{Calculating the mass transfer rate} \label{sec:heat_flux}

This section describes the calculation of the mass transfer rate, $\mdotS$, as given by \eqref{eq:discrete_mdot}. The numerical procedure calculates a value for $\mdotS$ in all cells containing a portion of the interface, known as mixed cells. We repeat \eqref{eq:discrete_mdot} here for some mixed cell $i,j,k$,
\begin{equation} \label{eq:mdot_calc}
\mdotS_{i,j,k} = \dfrac{1}{h_{fg}} \left( k_g \nabla_{n_\Gamma}^h T\vert^g_{i,j,k} - k_\ell \nabla_{n_\Gamma}^h T\vert^\ell_{i,j,k} \right) \,,
\end{equation}
where $\nabla_{n_\Gamma}^h$ indicates a discrete gradient either side of the interface. To calculate the normal temperature gradients, we disregard the temperature in mixed cells, since these temperatures essentially represents some volume average of the temperatures in both phases.

A method to calculate the interface normal temperature gradients was proposed by Kunkelmann \cite{Kunkelmann11} and also applied by Guedon \cite{GuedonThesis}. Temperature gradients are calculated in pure, single-phase cells which lie adjacent to mixed cells. Kunkelmann's method uses interpolation to obtain the $c=0.5$ iso-surface, the location of which is used to calculate a finite difference temperature gradient in these pure cells. The mass transfer rate in mixed cells are then obtained using an average from pure cell neighbours.

\begin{figure}
  \centering
  \begin{subfigure}[b]{0.52\textwidth}
    \def\svgwidth{0.9\textwidth}
    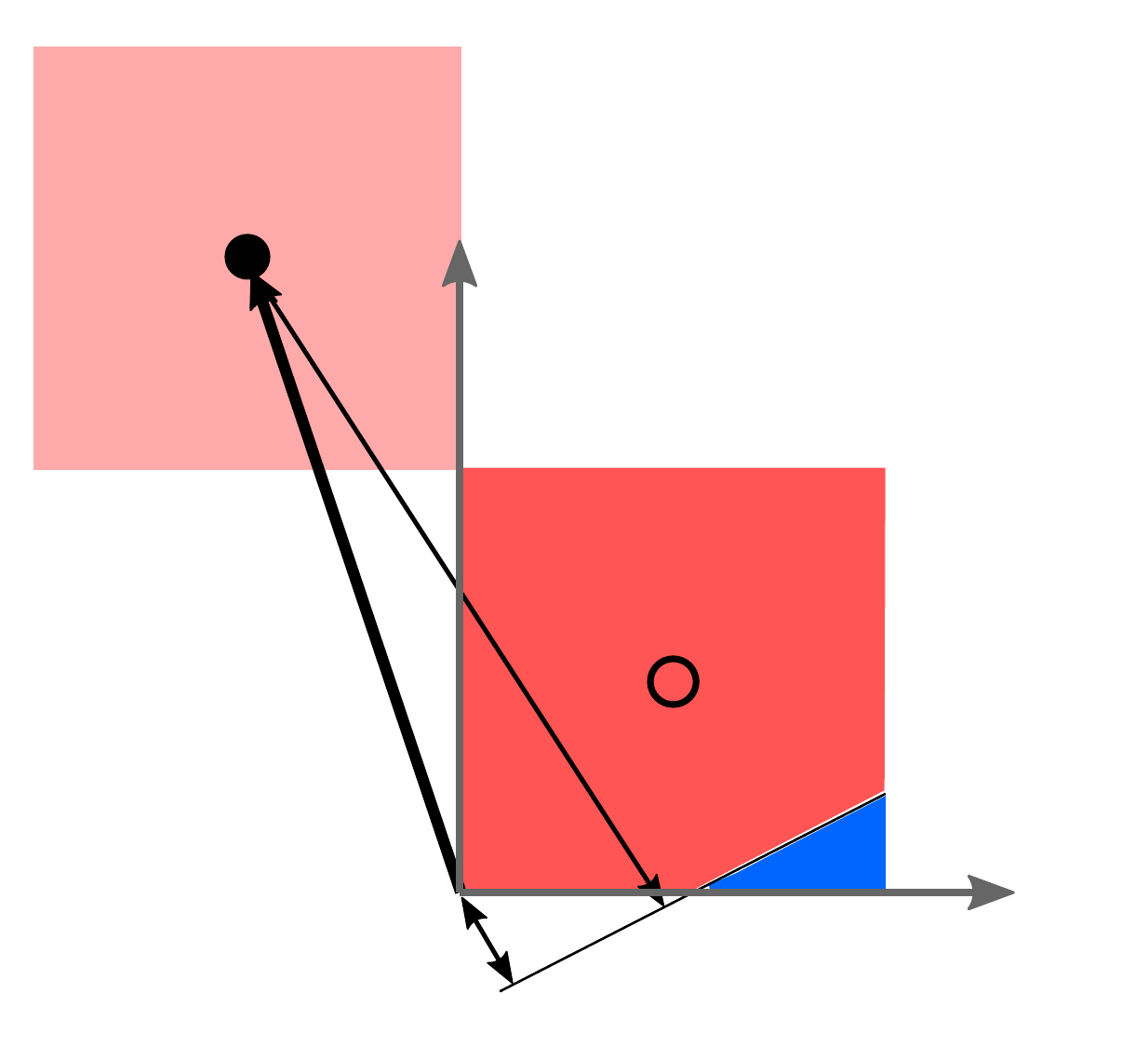
    \caption{Interface PLIC reconstruction in mixed cell $i,j,k$ with local coordinates $x_h$, $y_h$. $d_\Gamma$ is calculated in its pure cell neighbour.}
    \label{fig:dint_zoom}
  \end{subfigure}
  \hfill
  \begin{subfigure}[b]{0.45\textwidth}
    \def\svgwidth{0.9\textwidth}
\begingroup%
  \makeatletter%
  \providecommand\color[2][]{%
    \errmessage{(Inkscape) Color is used for the text in Inkscape, but the package 'color.sty' is not loaded}%
    \renewcommand\color[2][]{}%
  }%
  \providecommand\transparent[1]{%
    \errmessage{(Inkscape) Transparency is used (non-zero) for the text in Inkscape, but the package 'transparent.sty' is not loaded}%
    \renewcommand\transparent[1]{}%
  }%
  \providecommand\rotatebox[2]{#2}%
  \newcommand*\fsize{\dimexpr\f@size pt\relax}%
  \newcommand*\lineheight[1]{\fontsize{\fsize}{#1\fsize}\selectfont}%
  \ifx\svgwidth\undefined%
    \setlength{\unitlength}{690.77003002bp}%
    \ifx\svgscale\undefined%
      \relax%
    \else%
      \setlength{\unitlength}{\unitlength * \real{\svgscale}}%
    \fi%
  \else%
    \setlength{\unitlength}{\svgwidth}%
  \fi%
  \global\let\svgwidth\undefined%
  \global\let\svgscale\undefined%
  \makeatother%
  \begin{picture}(1,0.99162352)%
    \lineheight{1}%
    \setlength\tabcolsep{0pt}%
    \put(0,0){\includegraphics[width=\unitlength,page=1]{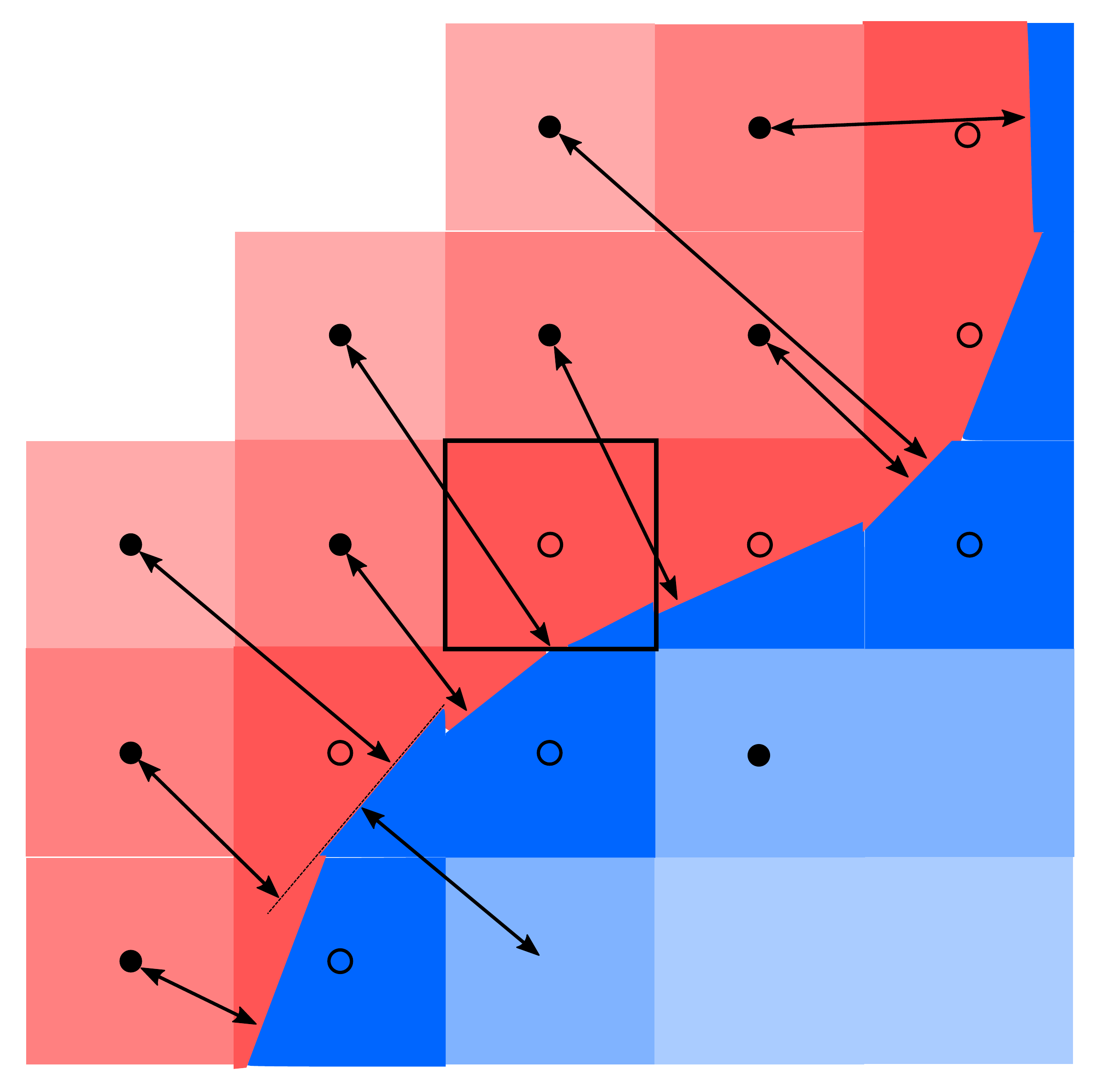}}%
    \put(0.41405334,0.54640719){\color[rgb]{0,0,0}\makebox(0,0)[lt]{\lineheight{0.625}\smash{\begin{tabular}[t]{l}$(i,j,k)$\end{tabular}}}}%
    \put(0.04642996,1.01105862){\color[rgb]{0,0,0}\makebox(0,0)[lt]{\begin{minipage}{0.93259992\unitlength}\raggedright \end{minipage}}}%
    \put(0,0){\includegraphics[width=\unitlength,page=2]{interface_heat_5x5.pdf}}%
  \end{picture}%
\endgroup%

    \caption{The mass transfer rate in mixed cell $i,j,k$ in the center, is calculated as a weighted average of pure cell neighbours in a $5 \times 5$ stencil around it.}
  	\label{fig:Tgrad_avg}
  \end{subfigure}
  \caption{Mass transfer rate calculation for mixed cells.}
  \label{fig:mdot_calc}
\end{figure}

We employ a similar approach to Kunkelmann, but instead of using an interpolation technique for the interface location, we use the PLIC reconstruction. The process is illustrated in Fig.~\ref{fig:mdot_calc}. Pure cells are indicated with filled dots, while mixed cells are indicated with open dots and contain a plane (PLIC) reconstruction of the interface. The temperature gradient normal to the interface is calculated in pure cells which are first (darker shade) or second (lighter shade) neighbours of mixed cells.

For a pure cell, the heat flux is simply estimated using a finite difference

\begin{equation}
k_p \nabla_{n_p}^h T = k_p \dfrac{T - T_\Gamma}{d_\Gamma} \, ,
\label{eq:grad_n_T}
\end{equation}
with $d_\Gamma$ the normal distance from the cell node to the interface and $p = \ell , g$ a phase indicator. It is assumed that $T_\Gamma=T_{sat}(p_{sys})$. The interface is defined in a mixed cell as a plane given by the implicit equation
\begin{equation}
\alpha= \bm{m} \cdot \bm{x} \, , 
\label{eq:plane_m}
\end{equation}
with $\alpha$ a constant and $\bm{m} = \langle m_1,m_2,m_3\rangle$ the interface vector. A local coordinate system is defined in each cell, with its origin in the corner closest to the larger domain origin. The interface vector is scaled such that 
\begin{equation}
\vert m_1 \vert + \vert m_2 \vert + \vert m_3 \vert = 1 \,,
\end{equation}
and is calculated using the VOF field with the Mixed Youngs-centered method \cite{Aulisa07}. The plane constant $\alpha$ is obtained by using geometrical volume calculations \cite{Scardovelli00}. In the local cell coordinate system of a mixed cell, the normal distance to the interface from the cell origin is therefore simply found by normalizing the interface vector \eqref{eq:plane_m}
\begin{equation}
d_{i,j,k}= \dfrac{\alpha}{\vert\vert \bm{m} \vert\vert} = \bm{n}_\Gamma^h \cdot \bm{x} \,,
\label{eq:cell_d}
\end{equation}
with $\bm{n}_\Gamma^h$ the planar interface normal vector.

To obtain the normal distance to the interface from any neighbouring pure cell, we apply a coordinate transformation. Let $ \Delta \bm{x}$ be the translation vector from the mixed cell origin to the pure cell node. The normal distance from this pure cell node is then 
\begin{equation}
d_\Gamma = d_{i,j,k} - \bm{n}_\Gamma^h \cdot \Delta \bm{x} \,.
\label{eq:dint_trans}
\end{equation}

Inside a pure cell, one can imagine that through the transformation in \eqref{eq:dint_trans}, the distance can be obtained to the interface representation in various neighbouring mixed cells. The interface distance is determined using the neighbouring mixed cell of which the collinearity of the plane normal in that cell and the translation vector is greatest, with the collinearity given by

\begin{equation}
\xi = \vert \bm{n}_\Gamma^h \cdot \Delta \bm{x} \vert \,.
\end{equation}

The interface heat flux can now be calculated in the pure cell using \eqref{eq:grad_n_T}, since its temperature is known and the interface normal distance is calculated, Fig.~\ref{fig:dint_zoom}. This is done for all pure cells that are first or second neighbours to mixed cells. Finally, the fluxes required to compute phase change in \eqref{eq:mdot_calc} are computed by using a weighted average of pure cell heat fluxes. This is done by considering all the first and second, pure cell neighbours in the desired phase within a $5 \times 5$ stencil, as shown in Fig.~\ref{fig:Tgrad_avg}. The example of liquid is given here, but the same applies for the gas phase:

\begin{equation} \label{eq:normTgrad}
\nabla_{n_\ell}^h T\vert^\ell_{i,j,k} = \sum\limits_{q=1}^n \chi_q \dfrac{T^{\ell}_q - T_\Gamma}{d_{\Gamma,q}} \,,
\end{equation}
where $q$ is an index for all $n$ neighbours in the stencil. The normalized weighting factor $\chi_q$ is given by

\begin{equation}
\chi_q = \dfrac{\gamma_q}{\sum\limits_{c=1}^n \gamma_c} \,,
\end{equation}
with $\gamma$ determined by the collinearity of the neighbour and the square of its distance from the mixed cell

\begin{equation}
\gamma_q = \dfrac{\xi_q}{\vert\vert \Delta \bm{x}_q \vert\vert ^2}
\end{equation}

The temperature gradients calculated in \eqref{eq:normTgrad} on either side of the interface are then used in \eqref{eq:mdot_calc} to obtain $\mdotS$ in all mixed cells.

\subsection{Geometric VOF advection for phase change} \label{sec:VOF_adv}
Several VOF advection schemes exist for incompressible, divergence free flow fields. For more details on geometric methods, refer to \cite{Scardovelli03},\cite[~p.95]{Book11}. The schemes available in \paris include the method of Li \cite{Li95} (sometimes referred to as \emph{CIAM} or Lagrangian Explicit) and the conservative method of Weymouth and Yue \cite{Weymouth10}. However, these methods cannot be applied directly to \eqref{eq:VOF_num}, due to the existence of the phase change source term.

In this work we present a novel method to solve \eqref{eq:VOF_num} using existing geometric advection methods designed for divergence-free velocity fields, while retaining the sharp interface approach. The method splits \eqref{eq:VOF_num} into two steps: 

\begin{equation}
\dfrac{\partial c}{\partial t} \approx \dfrac{c^{n+1} - c^n}{\Delta t} = \dfrac{1}{V} \int_V \dfrac{\partial H_1}{\partial t} + \dfrac{\partial H_2}{\partial t} \, dV \,,
\end{equation}
where the numeric subscripts indicate the step numbers.

The first step can be seen as a material advection of the liquid, in which the Heaviside function \eqref{eq:cons_Heavy} gets advected with the liquid velocity field. The second step accounts for the phase change term, or the relative velocity between the interface and liquid velocities. These steps are detailed next and are shown schematically in Fig.~\ref{fig:vof_adv}.

\subsubsection{Step 1: Advecting with an extended, divergence free liquid velocity}

The first step advects the Heaviside function with the liquid velocity $\bm{u}_\ell$, to produce an intermediate VOF field, $\tilde{c}$: 

\begin{equation}\label{eq:vof_step1_full}
\dfrac{\tilde{c} - c^n}{\Delta t} = \dfrac{1}{V} \int_V \dfrac{\partial H_1}{\partial t} \, dV = - \dfrac{1}{V} \oint_S H^n \bm{u}_\ell \cdot \bm{n} \, dA\,= - \nabla \cdot \left( \bm{u}_\ell c \right) .
\end{equation}

This step effectively assumes that the VOF field is advected at the velocity of the liquid phase, which is indeed the case everywhere inside the liquid. In order to use the existing geometric advection schemes, the velocity in the liquid is extended across the interface in a manner that ensures 

\begin{equation}
\bm{\nabla} \cdot \bm{u}_\ell = 0 \,.
\end{equation}

For this purpose we propose a simple yet effective technique to obtain $\bm{u}_\ell$, which is readily implementable into existing incompressible flow solvers. After solution of equations \eqref{eq:press_full_num} and \eqref{eq:correct_full_num}, the one--fluid velocity field ($\bm{u}^{n+1}$) is obtained, which will contain a discontinuity at the interface when phase change occurs (for phases with different densities).

\begin{figure}
  \centering
  \begin{subfigure}[b]{0.49\textwidth}
    \def\svgwidth{0.95\textwidth}
    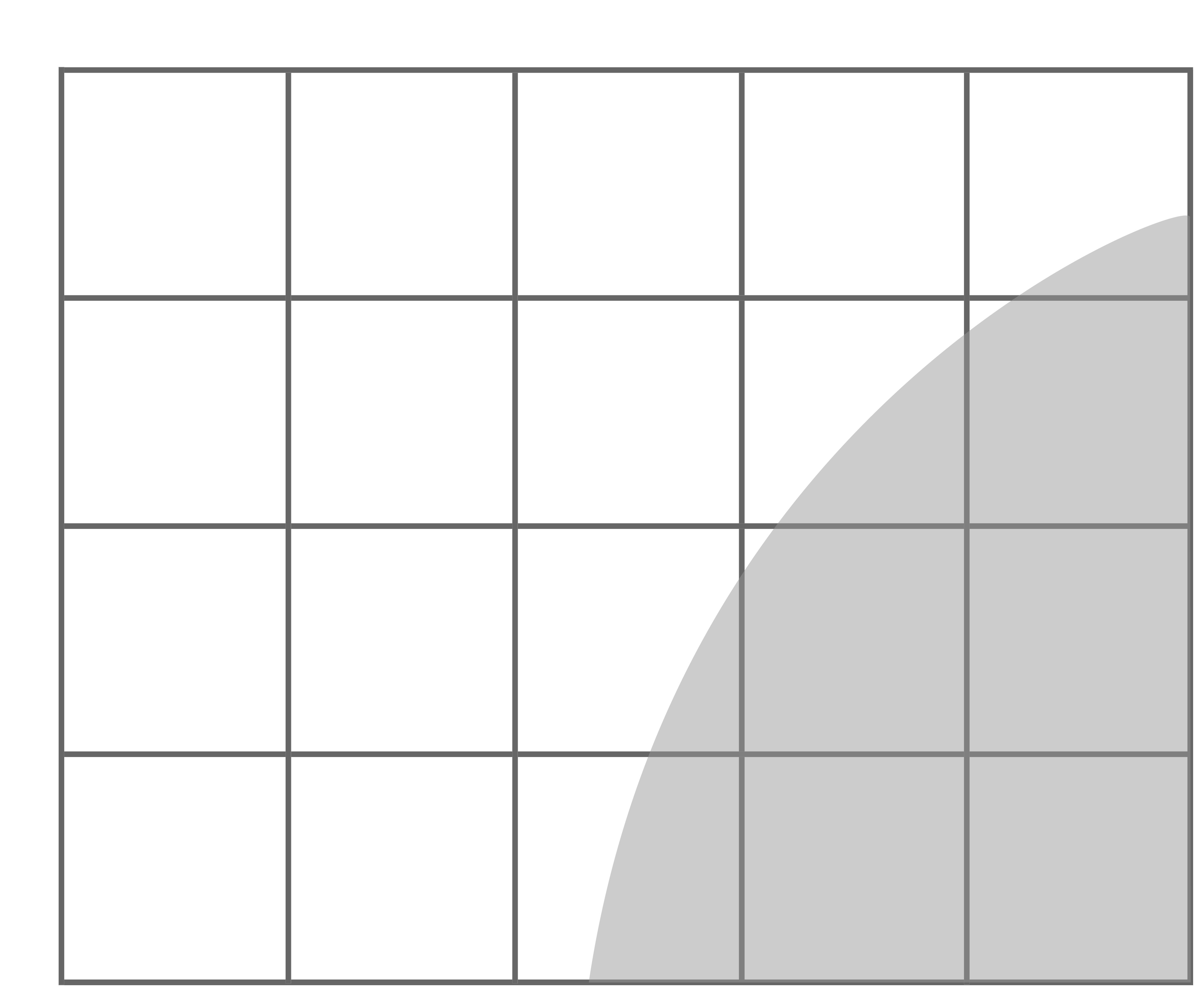
    \caption{$c^n$ and one--fluid velocity $\bm{u}$.}
    \label{fig:cn}
  \end{subfigure}
  \hfill
  \begin{subfigure}[b]{0.49\textwidth}
    \def\svgwidth{0.95\textwidth}
    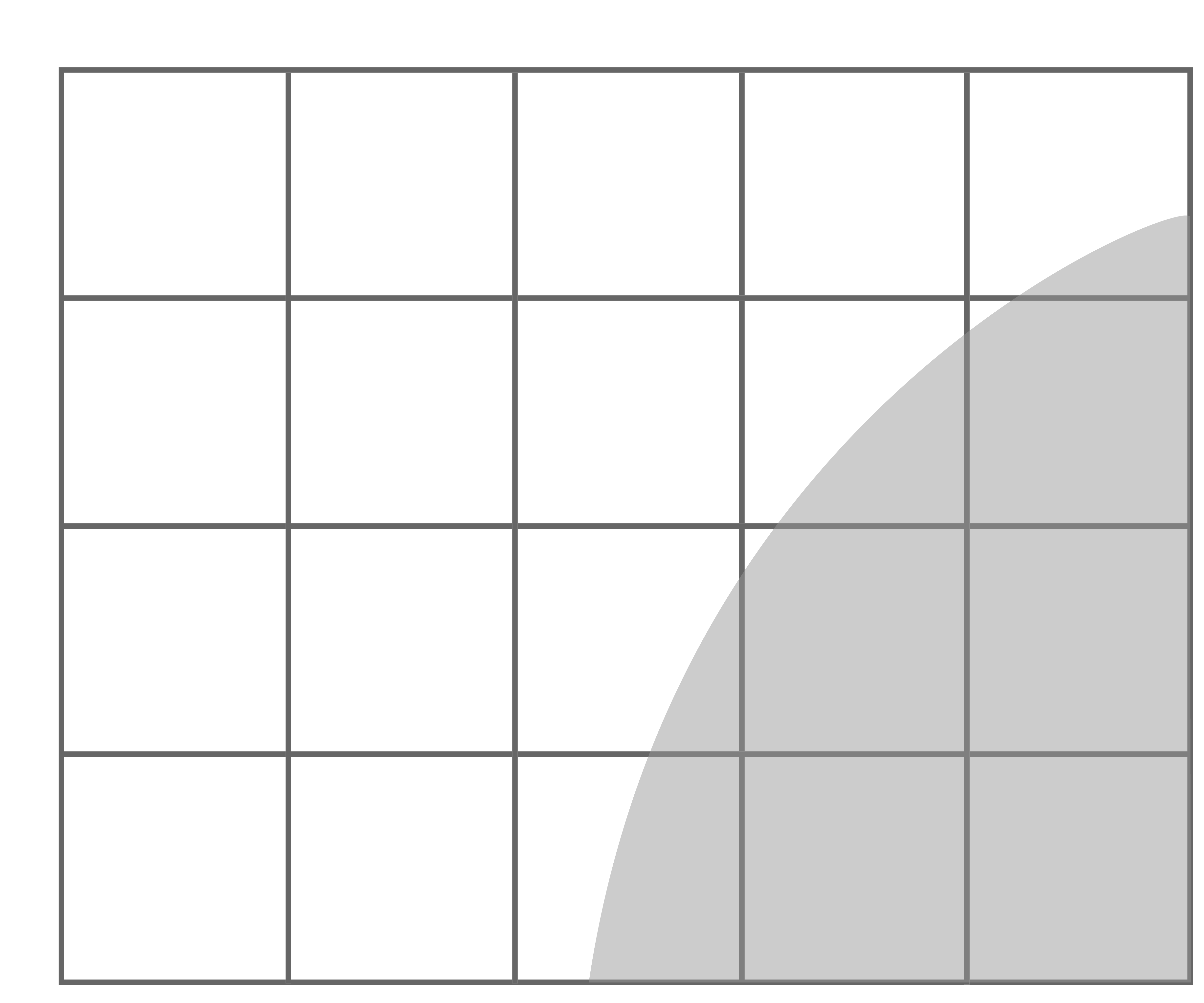
    \caption{Solution of $\tilde{\bm{u}}$ in a sub-domain.}
    \label{fig:us}
  \end{subfigure} \\
  \begin{subfigure}[b]{0.49\textwidth}
    \def\svgwidth{0.95\textwidth}
    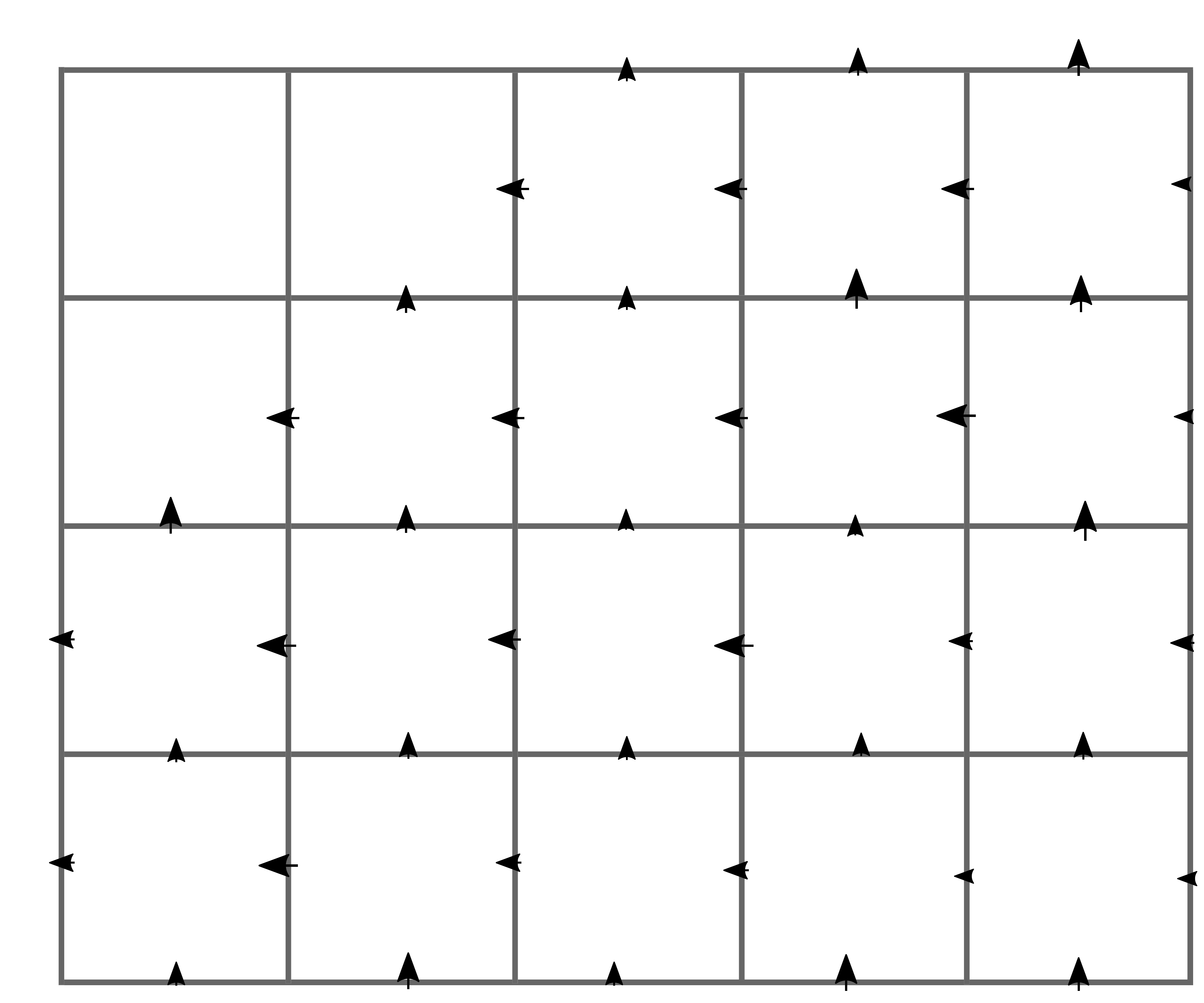
    \caption{Step 1 of VOF advection with divergence free $\bm{u}_\ell$ to obtain $\tilde{c}$.}
    \label{fig:c_tilde}
  \end{subfigure}
  \hfill
  \begin{subfigure}[b]{0.49\textwidth}
    \def\svgwidth{0.95\textwidth}
    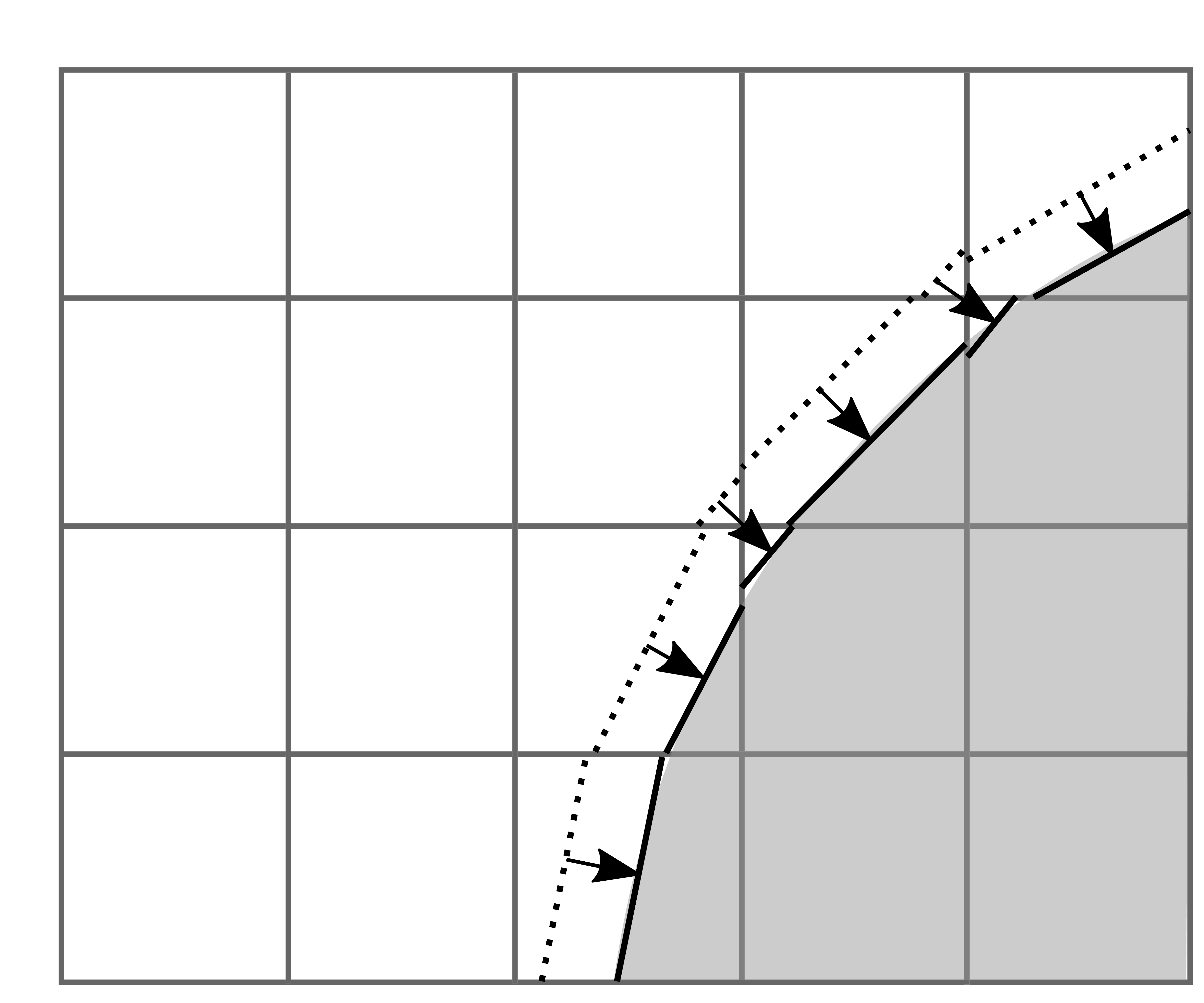
    \caption{Step 2: Interface shift to account for phase change.}
    \label{fig:c_n+1}
  \end{subfigure}
  \caption{Schematic of split geometric VOF advection method for an evaporating droplet.}
  \label{fig:vof_adv}
\end{figure}

We now construct a sub-domain around the interface, of which a 2D example is shown in Fig.~\ref{fig:us}. The sub-domain is delimited on the liquid side of the interface by defining all the cell faces between mixed and liquid cells as domain boundaries on which a no-slip boundary condition is applied (red line). On the other side of the interface, all the cell faces between second and third pure gas cell neighbours (blue dotted line) are considered as free-flow faces (a fixed pressure is applied with a zero velocity gradient normal to the boundary).

A Poisson problem is set up on this newly created sub-domain in a manner similar to \eqref{eq:press_full_num}

\begin{equation}
\bm{\nabla}^{h} \cdot \left[\frac{\Delta t}{\rho^{n+1}} \bm{\nabla}^{h} \tilde{p} \right] = \mdotS \left( \frac{1}{ \rho_g} - \frac{1}{\rho_\ell} \right) \delta_\Gamma \,,
\label{eq:press_inv}
\end{equation}
with the same phase change source term, but of opposite sign. This equation is solved with the same convergence tolerance as for the previous problem. A phase change related velocity correction is then obtained on the cell faces of the newly defined sub-domain, similar to \eqref{eq:correct_full_num}

\begin{equation}
\tilde{\bm{u}} = -\dfrac{\Delta t}{\rho^{n+1}}\bm{\nabla}^{h} \tilde{p}^{n+1} \,.
\end{equation}
with $\tilde{\bm{u}}$ the newly obtained velocity field. It is defined everywhere on the cell faces inside the light blue sub-domain in Fig.~\ref{fig:us}, elsewhere it is zero. The divergence free liquid velocity field can now be found using

\begin{equation}
\bm{u}_\ell^{n+1} = \bm{u}^{n+1} + \tilde{\bm{u}} \,.
\end{equation}
This step extends the velocity of the liquid phase across the interface for all faces up to the third neighbours in the gas phase. Velocities that are further into the gas phase are not of concern, as they do not affect the VOF field. The $\bm{u}_\ell$ field is divergence free inside the liquid and up to the extension into the gas phase by construction
\begin{equation}
\bm{\nabla} \cdot \bm{u}_{\ell} = \bm{\nabla} \cdot \bm{u}^{n+1} + \bm{\nabla} \cdot \tilde{\bm{u}} = \mdotS \left( \frac{1}{ \rho_g} - \frac{1}{\rho_\ell} \right) \delta_\Gamma - \mdotS \left( \frac{1}{ \rho_g} - \frac{1}{\rho_\ell} \right) \delta_\Gamma = 0 \,.
\end{equation}

We now use existing geometric advection schemes to calculate $\tilde{c}$, shown in Fig.~\ref{fig:c_tilde}.

\subsubsection{Step 2: Shifting the interface to account for phase change}

In the second step of the split VOF equation, the interface shift due to phase change is accounted for
\begin{equation} \label{eq:vof_step2_full}
\dfrac{c^{n+1} - \tilde{c}}{\Delta t} = \dfrac{1}{V}  \int_V \dfrac{\partial H_2}{\partial t} \, dV = - \dfrac{1}{V} \int_\Gamma \left(\bm{u}_\ell - \bm{u}_{\Gamma} \right) \cdot \bm{n}_{\Gamma}dA = - \dfrac{\mdotS}{\rho_{\ell}} \, .
\end{equation}

This is illustrated in Fig.~\ref{fig:c_n+1}. This step accounts for the relative movement between the interface and neighbouring liquid molecules (bearing in mind that the liquid is the tracked phase). The relative velocity $(\bm{u}_\Gamma - \bm{u}_\ell) \cdot \bm{n}_\Gamma$ is a consequence of mass transfer, \eqref{eq:mdot}.

To satisfy \eqref{eq:vof_step2_full}, we need to shift the interface in the normal direction. We define the shift in the interface $\Delta d$

\begin{equation}
\Delta d = (\bm{u}_\Gamma - \bm{u}_\ell) \cdot \bm{n}_\Gamma = - \dfrac{\mdotS}{\rho_{\ell}} \dfrac{\Delta t}{h} \,,
\end{equation}
with $h$ the cell length used to re-scale the interface shift to the local cell coordinate system (see Fig.~\ref{fig:dint_zoom}), wherein the PLIC plane constant is defined \eqref{eq:plane_m}. We once again use the PLIC reconstruction and existing geometric VOF library in \paris \cite{Scardovelli00} to calculate $c^{n+1}$

\begin{align}
\alpha^{n+1} &=\tilde{\alpha} + \vert\vert \bm{m} \vert\vert \Delta d \ldots 0 \leq \alpha^{n+1} \leq 1 \notag \\
c^{n+1} &= \ f(\alpha^{n+1},\bm{m}) \,.
\end{align}
The movement is capped to avoid local over- and undershoots and respect $0 \leq c^{n+1} \leq 1$. When they do occur, the clipped amount is accounted for in the neighbour that is located in the direction of interface movement in order to still respect mass conservation. 

When this step is completed, the VOF equation \eqref{eq:VOF_num} is satisfied. The advection of thermal energy is solved such that fluxes in heat capacity are calculated in a consistent manner with the VOF fluxes, as detailed next.

\subsection{Energy conservation}
A thermal energy conservation equation \eqref{eq:et_disc} was added to \paris to facilitate the simulation of phase change flows.

\subsubsection{Calculating the energy advection term} \label{sec:et_adv}
The advection term is calculated as

\begin{equation}
C_p^\ast T^\ast - C_p^{n} T^{n} =  - \Delta t \sum\limits_{f} \, C_p^{n}\vert_f \, T^n_f u_f^n A_f \,,
\label{eq:disc_et_adv}
\end{equation}
where $C_p^\ast T^\ast$ indicates the thermal energy per unit volume after advection is considered. The right-hand-side represents the sum of thermal energy fluxes at all $f$ cell faces for time step $n+1$. Each face has an area $A_f=h^2$ (in 3D) and face velocity $u_f = \bm{u} \cdot \bm{n}_f$, which is calculated explicitly on the staggered (MAC) grid. Care needs to be taken when calculating $C_p^{n}\vert_f$ as well as $T^n_f$, respectively the fluxed heat capacity and face temperature. One reason is that the volumetric heat capacity ratio becomes significant in many problems of interest. Consider, as an example, the properties of water at atmospheric conditions in Table \ref{table:water_props}. In this case $\rfrac{C_{p,\ell}}{C_{p,g}} \approx 3200$, which motivates why a conservative formulation is preferable, as opposed to the non-conservative approximation

\begin{equation}
C_p^\ast T^\ast - C_p^n T^n \vert_{i,j,k} =  - \Delta t \, C_p^n \vert_{i,j,k} \, \sum\limits_{f} \, T^n_f u_f^n A_f \,.
\label{eq:disc_et_adv_noncons}
\end{equation}

\begin{table}
	\centering
		\begin{tabular}[width=0.75\textwidth]{lccc}
		\hline
		\textbf{Phase} & \textbf{Density}  & \textbf{Specific heat capacity} & \textbf{Vol. heat capacity} \\ 
		$ $ & $\rho$ $\left[kg.m^3\right]$ & $c_p$ $\left[kJ.kg^{-1}.K^{-1}\right]$ & $C_p$ $\left[kJ.m^{-3}.K^{-1} \right]$ \\
		\hline
		Gas (vapour) & $0.6$ & $2.080$ & $1.248$\\ 
		\hline
		Liquid & $958$ & $4.216$ & $4.039$ $\times 10^3$ \\ 
		\hline
		\end{tabular}
		\caption{Properties for saturated water at atmospheric conditions}
		\label{table:water_props}
\end{table}

A flux-consistent, geometric advection technique is employed. The term flux-consistent is used, since the flux terms in \eqref{eq:disc_et_adv} are calculated using the same geometric advection procedure as for the VOF function. Note that the flux-consistent implementation here is inspired by a similar method already present in \textsc{PARIS}, but applied to the momentum term \cite{Fuster18}.

In the same manner as the directionally split VOF advection methods (\cite{Li95, Weymouth10}) three separate, consecutive sweeps are performed, one for each coordinate direction (in 3D). Let $s$ be the sweep counter, so that $s=1$ corresponds to time step $n$, $s=1,4$ and $s=4$ corresponds to time step $n+1$. Before the sweeps start, a thermal energy $e_t$ is calculated for every cell
\begin{equation}
e_t^{(s=1)} = e_t^n = C_p^n T^n \,,
\end{equation}
where the cell temperature is known from the solution at time step $n$ and the volumetric heat capacity in a cell is obtained by using a volume average of the phase heat capacities
\begin{equation}
C_p^{(s=1)} = C_p^n = C_{p,\ell} \, c + C_{p,g} \left( 1 - c \right) \,.
\end{equation}

For each direction sweep, the following process is followed:

\begin{enumerate}
\item The face temperatures $T_f$ are obtained by using a fifth order WENO reconstruction scheme \cite{Shu09}.
\item The fluxed volumes and their respective VOF values are known from the geometric VOF flux calculations at each of the two faces in the specific sweep direction. This allows us to calculate a volume weighted volumetric heat capacity $C_p\vert_f$ and the fluxed thermal energy at each face from the temperature in the WENO scheme $C_p\vert_f\,T_f$. We now have the intermediate thermal energy in the cell (after the sweep) and can calculate the intermediate temperature

\begin{equation}
T^{(s)} = \dfrac{e_t^{(s)}}{C_p^{(s)}} \,,
\end{equation}
\item The thermal energy and VOF sweeps are done in tandem for every sweep direction. Before the next direction is swept, the interface is reconstructed from the newly calculated intermediate VOF $c^{(s+1)}$, along with the thermal energy at the same sweep $e_t^{(s+1)}$. This ensures that the heat capacity that is used at each fluxing step is consistent with the VOF function. The face temperatures are calculated by using the updated temperature field after each sweep.
\end{enumerate} 

After the three coordinate directions have been swept using $\bm{u}_\ell$, (just like the VOF process) the energy change due to phase change is calculated by 

\begin{equation}
\Delta e_t = \Delta c \left( C_{p,\ell} - C_{p,g}\right) T_\Gamma
\end{equation}

\subsubsection{Calculating the energy diffusion term} \label{sec:et_diff}
The heat diffusion term needs to be evaluated to complete the integration of \eqref{eq:et_disc} in time. This term is calculated implicitly using a similar approach to Sato and Ni\v{c}eno \cite{Sato13}. The saturation temperature at the interface is applied directly by solving the diffusion term separately for each phase. The interface is treated as a boundary, where a Dirichlet boundary condition is specified for the temperature. This is equivalent to solving two heat diffusion equations with constant fluid properties, with the interface an arbitrary boundary separating the two phases in which each respective problem is solved.

\begin{figure}[ht]
	\centering
	\def\svgwidth{0.85\textwidth}
	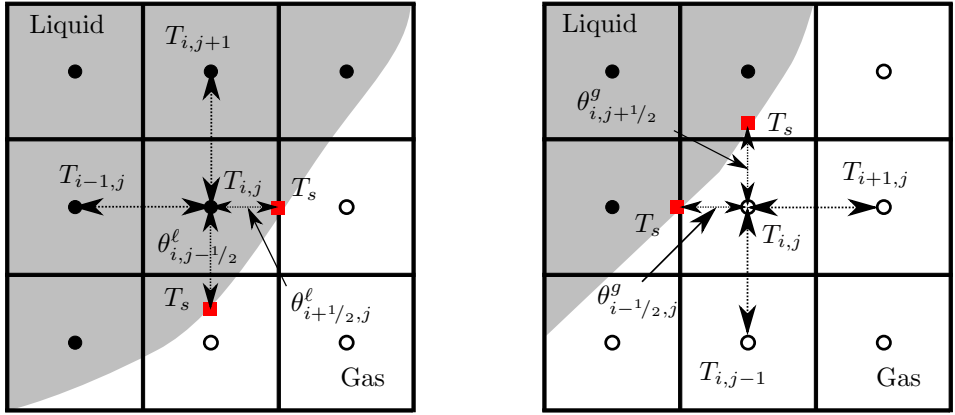
	\caption{Discretisation for temperature neighbouring the interface. On the left the computational stencil for a liquid cell is shown, with the stencil for its neighbour in the gas on the right.}
	\label{fig:T_diff_stencil}
\end{figure}

First, the phase of a specific node is determined by simply evaluating the VOF function: A cell node is determined to be inside the liquid whenever $c \geq 0.5$, otherwise it is in the gas. Fig.~\ref{fig:T_diff_stencil} shows two 2D stencils used to compute the diffusion term for a liquid (left hand side) and gas (right hand side) node. 

Consider the liquid node on the left as an example. The finite volume discretiation of the diffusion term for the 2D cell at position $i,j$ is approximated using the irregular stencil as
 
\begin{equation}
\dfrac{1}{V}\int_{V_{i,j}} \bm{\nabla} \cdot 
\left( k \bm{\nabla} T \right) dV \approx k_{\ell} \left( \dfrac{\nabla^{h}_y \,T_{i,j+\rfrac{1}{2}} 
- \nabla^{h}_y \,T_{i,j-\rfrac{1}{2}}}{\rfrac{1}{2} \left( h + \theta^{\ell}_{i,j-\rfrac{1}{2}} \right) } + \dfrac{\nabla^{h}_x \, 
T_{i+\rfrac{1}{2},j} - \nabla^{h}_x \,T_{i-\rfrac{1}{2},j}}{ \rfrac{1}{2} \left( \theta^{\ell}_{i+\rfrac{1}{2},j} + h \right) } \right) \,, 
\label{T_diff_disc}
\end{equation}
noting that $\Delta x = \Delta y = h$ and $k=k_{\ell}$ since we are only considering the liquid.

The temperature gradient is approximated using a finite difference. Consider again the liquid node:
\begin{equation} \label{eq:mod_branchesT}
\bm{\nabla}^{h}_x \,T_{i+\rfrac{1}{2},j} = \dfrac{T_\Gamma -
T_{i,j}}{\theta^{\ell}_{i+\rfrac{1}{2},j}} \,\,;\quad
\bm{\nabla}^{h}_x \,T_{i-\rfrac{1}{2},j} = \dfrac{T_{i,j} -
T_{i-1,j}}{h} \, .
\end{equation}
For the temperature gradient to the right of the node in question, the interface distance $\theta^{\ell}$ is used to calculate the finite difference. A standard finite difference is used when the temperature gradient is approximated between two nodes in the same phase.

The distance to the interface from a node is simply taken as the height function in that direction. When the interface configuration is such that a height cannot be obtained in the required direction, the distance is approximated by using a plane reconstruction of the interface in the staggered volume. The distance is capped so that $\epsilon h \leq \theta \leq h$, with a value of $\epsilon = 10^{-3}$ typically used.

\subsection{A note on Momentum Conservation}

Before proceeding to present results, a word is due on the numerical treatment of the momentum conservation equation. For the purpose of this work a non-conservative momentum advection formulation was used. The two-step advection process was not applied to the existing momentum conserving scheme \cite{Fuster18} in \paris due to additional numerical routines required to do so. One challenge is the fact that the momentum control volume is staggered to the pressure and thermal energy control volumes on the staggered (MAC) grid.

The error introduced with the use of the non-conservative discretization of the momentum equation manifests in the force balance on the interface. Without a conservative discretisation, an error is introduced for the vapour recoil pressure. This error was not found to be detrimental for the purpose of this work, but a conservative, consistent implementation for the momentum is envisaged in future work.

\section{Simulation Cases} \label{sec:results}

In this section we validate the novel numerical scheme with analytical benchmark test-cases. The numerical method was implemented in \paris with full parallel processing capability in three spatial dimensions using the MPI libraries \cite{MPI04}.

\subsection{Interface distance calculation}

This test case measures the accuracy of the PLIC--based interface distance calculation, as explained in Section \ref{sec:heat_flux}. It is clear from \eqref{eq:grad_n_T} that the interface distance calculation is important when the normal temperature gradients are calculated for the mass transfer rate. As a first validation test, the VOF field was initialized as a plane in 3D. The theoretical interface distances were compared to the calculated distances in \paris for all first and second pure cell neighbours. The calculated distances were found to correspond to the theoretical ones to machine precision (as expected).

Next, a 3D droplet test was performed. A droplet of radius $R_0=0.25+\delta_R$ is initialized at $x_0+\delta_x, \, y_0+\delta_y, \, z_0+\delta_z$ of a unit cube, with $x_0=y_0=z_0=0.5$. The $\delta$-values are each a randomly generated number such that $-h \leq \delta \leq h$, with $h=\rfrac{1}{N}$ the cell size for a mesh with $N^3$ grid points. Five different resolutions were tested: $N=\left\lbrace16,\,32,\,64,\,128,\,256\right\rbrace$. At each mesh resolution, $100$ different droplets were generated for a total of $500$ tests. 

For each droplet, the calculated interface normal distances $d^h$ were calculated for all first and second pure cell neighbours (in each phase). The relative error for a pure cell is calculated as  

\begin{equation}
e_{i,j,k} = \dfrac{\vert d_{i,j,k}^h - d \vert}{R_0} \,.
\end{equation}

For each droplet, the max error is recorded as well as an $L_2$-norm of all $n$ first- and second pure cell neighbours to mixed cells, with

\begin{equation}
L_2 = \sqrt{\dfrac{\sum_q^n e_q^2}{n}} \,.
\end{equation}

\begin{figure}[ht]
\centering
\includegraphics[width=.85\textwidth]{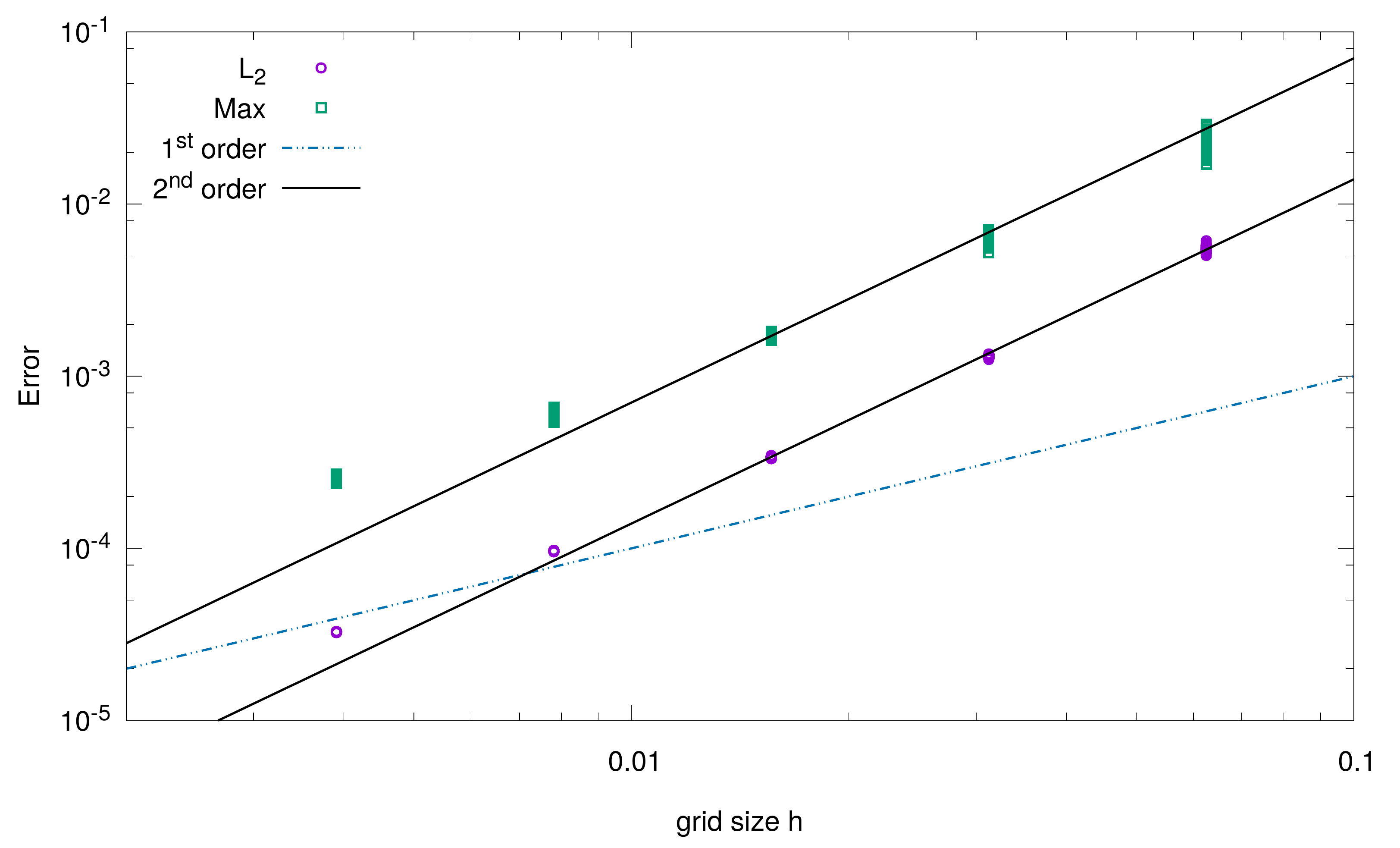} 
\caption{Convergence of the PLIC interface distance calculation.}
\label{fig:eDist}
\end{figure}

The results are presented in Fig.~\ref{fig:eDist}. The convergence rate for the ($L_2$) average distance error is around second order at lower mesh resolutions. For higher mesh resolutions it is between first and second order.

\subsection{Two-dimensional droplet evaporating test}
One crucial aspect of the novel numerical method is the proposed two-step VOF advection procedure. This includes the correct calculation of a divergence free liquid velocity and subsequent phase change adjustment to ensure mass conservation. A two-dimensional evaporating droplet is modelled to validate this part of the implementation. The discretization errors that may be present from the mass transfer rate calculations are removed by specifying a constant mass transfer rate of $\mdotS=0.05$, ensuring that this test specifically evaluates the two-step VOF-advection procedure.

A two-dimensional, circular liquid droplet with initial radius $R=0.23$ is initialized in the center of a square domain of unit length. The units in this problem are irrelevant and we can consider the problem dimensionless. The boundaries are given an outflow boundary condition (fixed pressure and zero normal gradient for the velocity). The constant mass transfer rate causes the droplet to evaporate. The radius evolution over time is simply

\begin{equation}
R \left( t \right) = R_0 - \mdotS t \,.
\end{equation} 

\begin{figure}[ht]
\centering
\includegraphics[width=.55\textwidth]{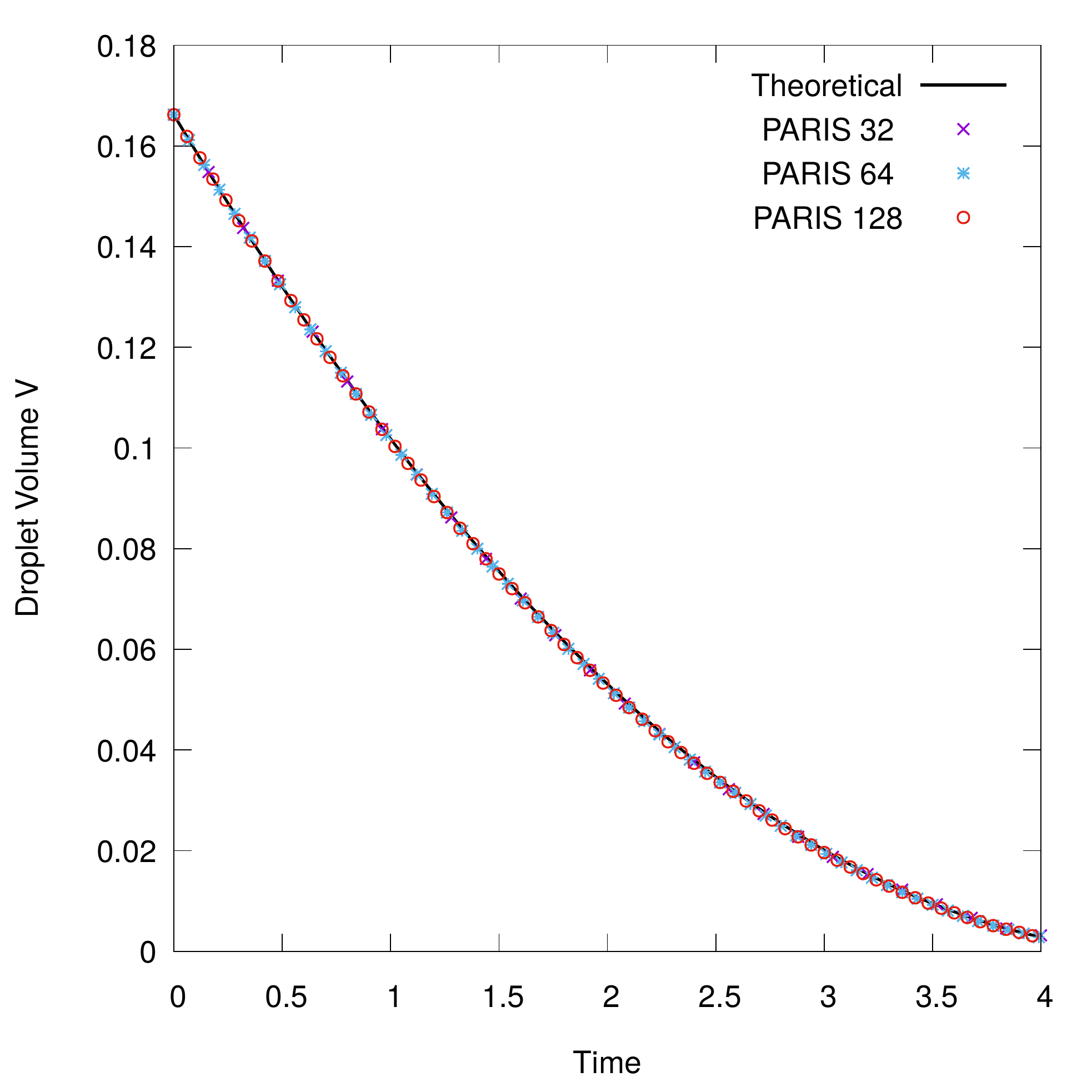} 
\caption{The volume evolution an evaporating cylinder in 2D with constant rate of evaporation.}
\label{fig:evap_drop_V}
\end{figure}

\begin{figure}[!ht]
\centering
\begin{tabular}[width=1.00\textwidth]{cc}
\includegraphics[height=40mm]{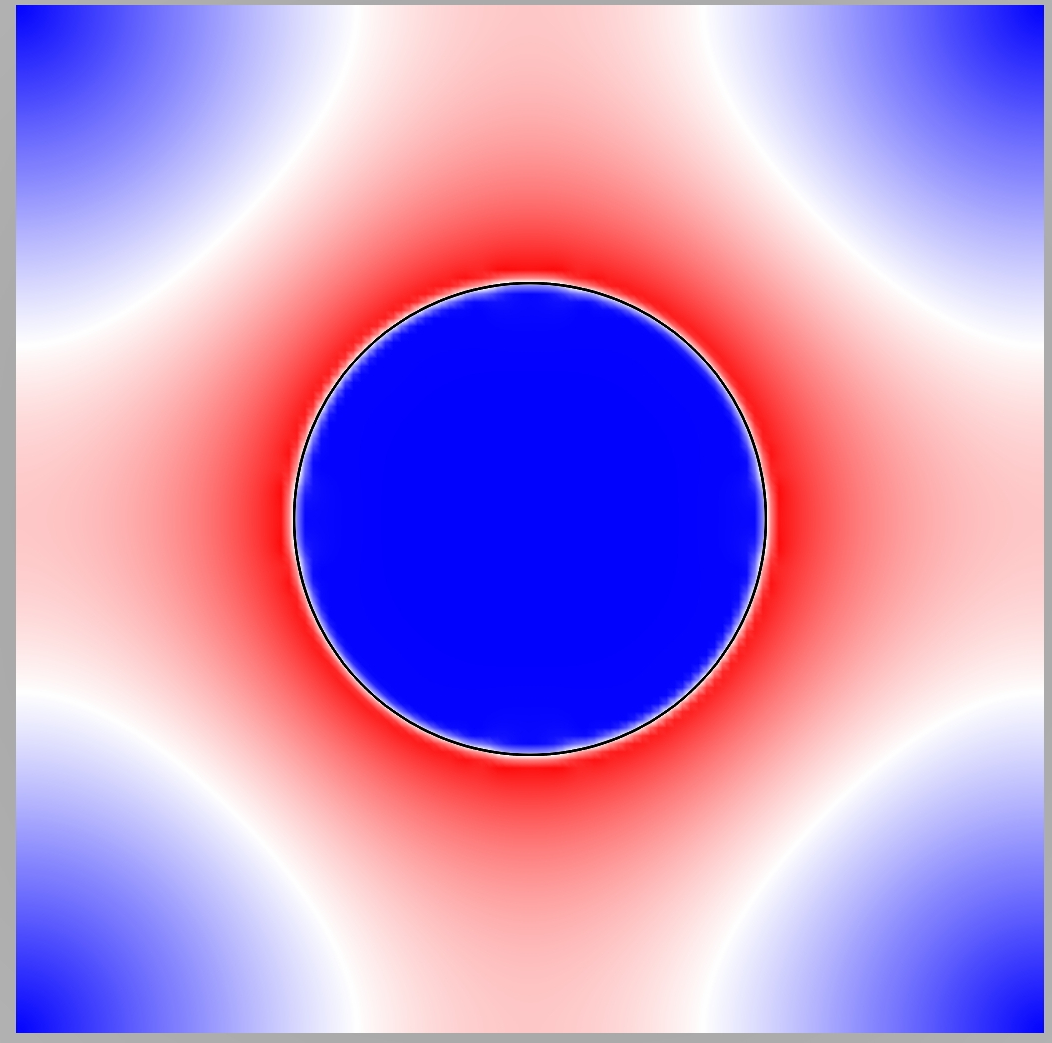} & 
\includegraphics[height=40mm]{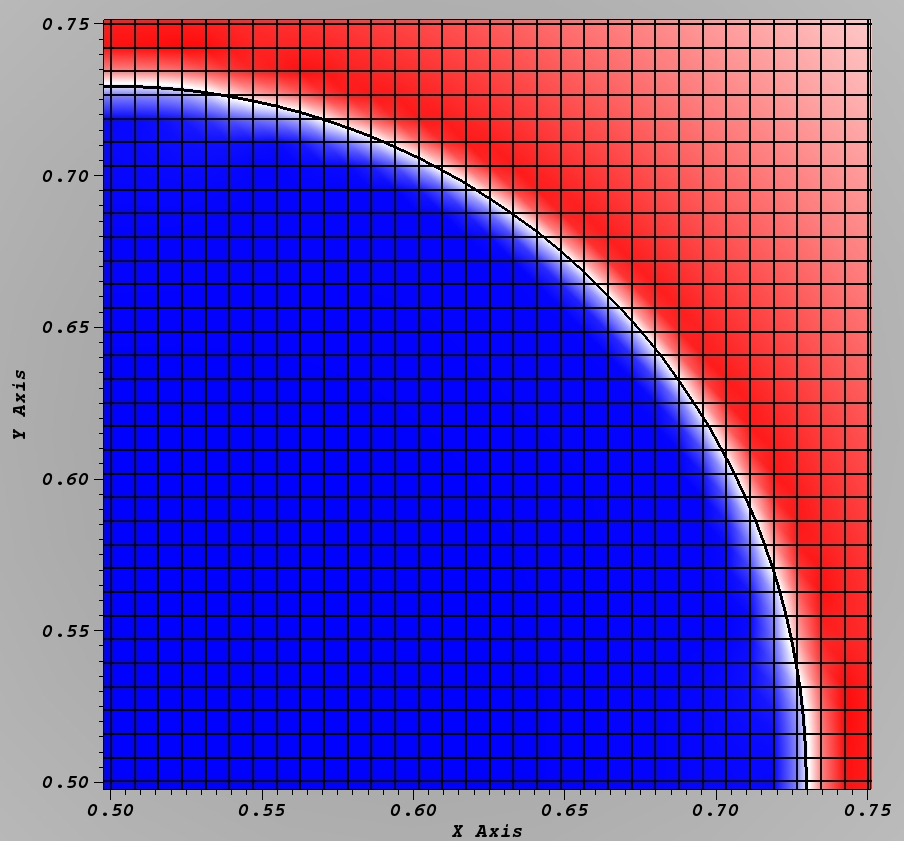} \\
\includegraphics[height=40mm]{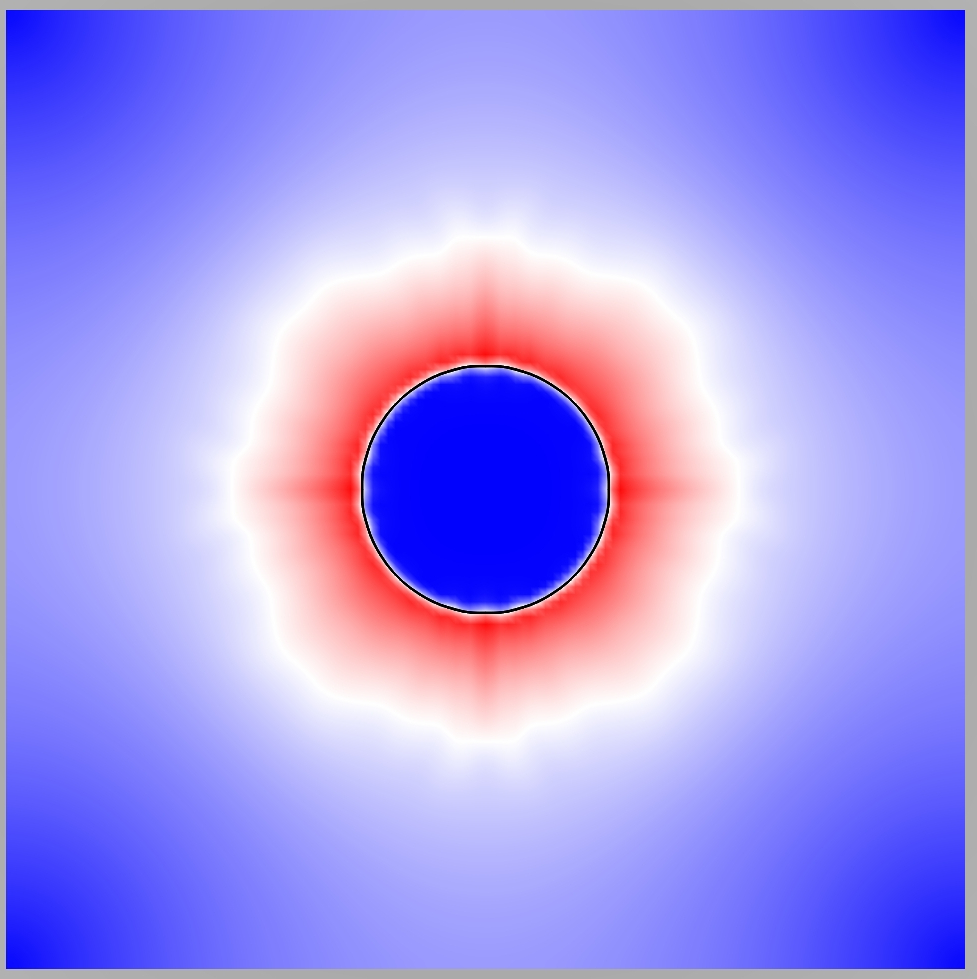} & 
\includegraphics[height=40mm]{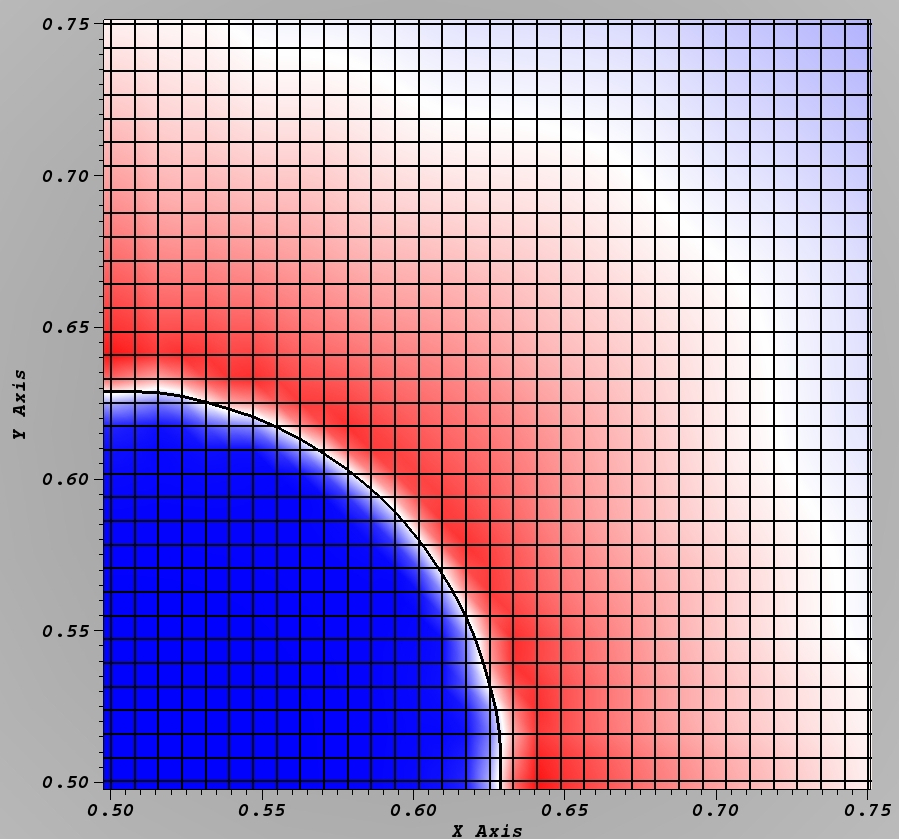} \\
\includegraphics[height=40mm]{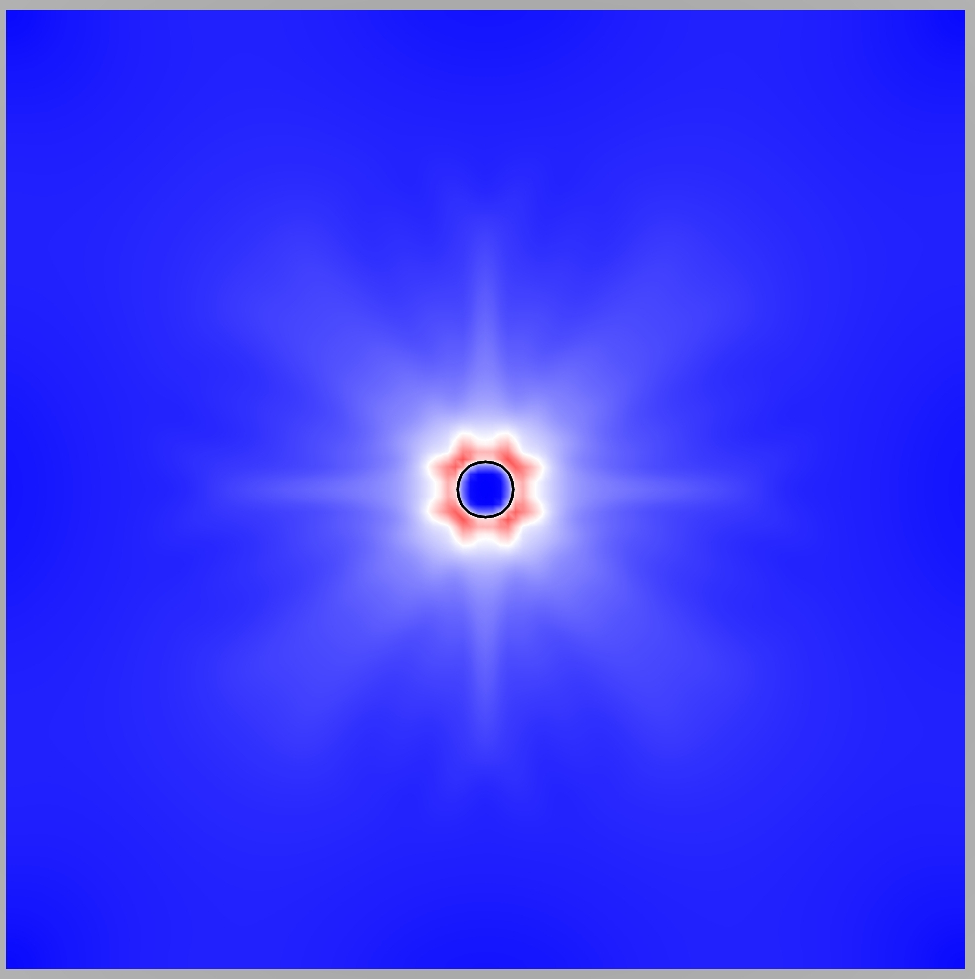} & 
\includegraphics[height=40mm]{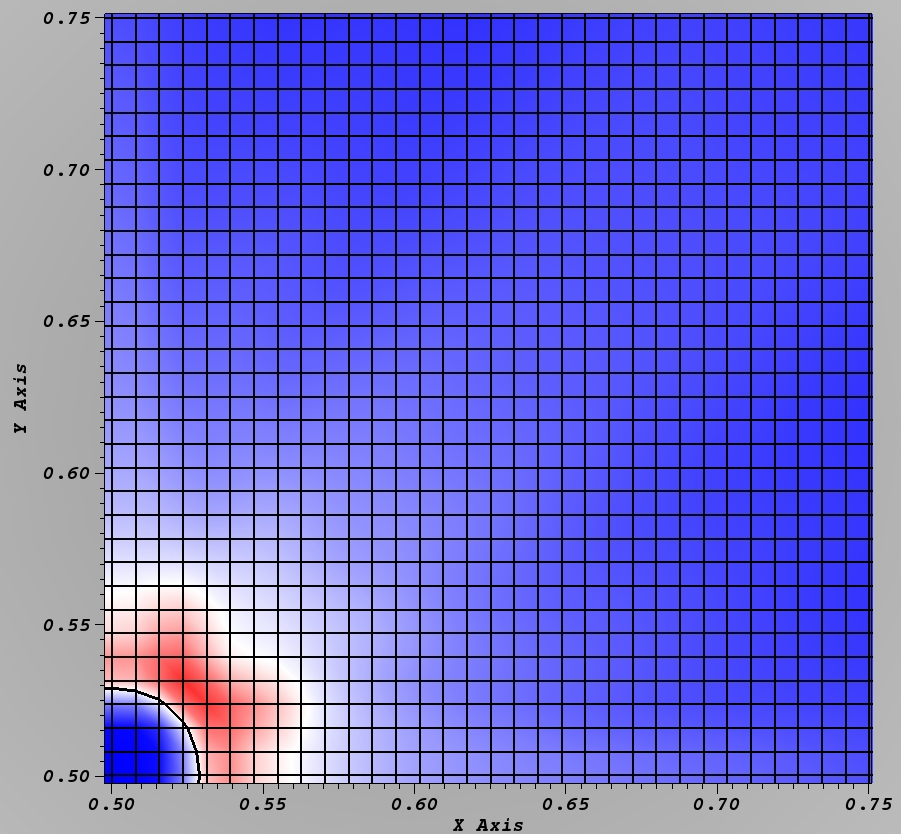} \\
\end{tabular}
\caption{A time series showing the velocity magnitude for the finest resolution simulation ($N_x=128$), taken at $0.01$, $2.01$ and $4.01$s. A zoom showing the mesh is on the right.}
\label{fig:vel_series}
\end{figure}

Three grid resolutions were used: $N_x=32,64$ and $128$ with respective time step sizes of $t=0.002\,s$, $t=0.001\,s$ and $t=0.0005\,s$. The time evolution of the droplet volume is shown in Fig.~\ref{fig:evap_drop_V} and compared to the analytical solution. As shown, an accurate solution is achieved on all meshes. Fig.~\ref{fig:vel_series} shows a time sequence of the droplet evolution for the finest resolution. The velocity magnitude is shown, revealing resulting radial flow patterns. It is encouraging to note that the circular shape is maintained throughout and the velocity field around the droplet remains symmetric.

%

\subsection{Stefan problem}
The Stefan problem was used by Welch and Wilson \cite{Welch00} to validate their VOF-based method and has since often been used to validate incompressible phase change methods \cite{GuedonThesis, Sato13}. In this problem, a liquid at saturation temperature is initially at rest. A thin vapour layer is present between the liquid and a wall, which is at a fixed, elevated temperature of $T_{sup}$ degrees above the saturation temperature, $T_{wall} = T_{sat} + T_{sup}$. Gravity is neglected and an outflow boundary condition is used opposite the heated wall.

The temperature gradient between the wall and the interface, which is at saturated temperature, causes phase change and the vapour layer to grow in time. The liquid is pushed out of the domain by the growing vapour layer. Let the thickness of the vapour layer be $\delta$. An analytical solution is available for the evolution of the interface thickness and temperature profile \cite{Welch00}. The interface thickness is given by

\begin{equation}
\delta(t) = 2 \lambda \sqrt{\alpha t} \,,
\label{eq:stefan_delta}
\end{equation}
with $t$ the time and $\alpha=\rfrac{k_g}{\rho_g c_{p,g}}$ the thermal diffusivity of the vapour phase. $\lambda$ is found by solving the transcendental equation

\begin{equation}
\lambda \exp( \lambda^2)erf(\lambda) = \dfrac{c_{p,g}T_{sup}}{h_{fg}\sqrt{\pi}} \,.
\label{eq:stefan_lamda}
\end{equation}

A two-dimensional test case was defined with a domain size of $10 \, mm$, a $T_{sup} = 10 \,K$ superheat and fluid properties for water at atmospheric conditions. These properties are given in Table~\ref{table:atm_water}.

\begin{table}
	\centering
		\begin{tabular}[width=0.85\textwidth]{lccc}
		\hline
		\textbf{Property} & \textbf{Units} & \textbf{Liquid}  & \textbf{Vapour} \\ 
		\hline
		Density $\rho$ & $\left[kg.m^3\right]$ & $958$ & $0.6$ \\ 
		\hline
		Specific heat $c_p$ & $\left[kJ.kg^{-1}.K^{-1}\right] $ & $4216$ & $2080$\\ 
		\hline
		Viscosity $\mu$ & $[Pa.s]$ & $2.82 \times 10^{-4}$ & $1.23 \times 10^{-5}$ \\ 
		\hline
		Thermal conductivity $k$ & $[W.m^{-1}.K^{-1}]$ & $0.68$ & $0.025$ \\ 
		\hline
		Surface tension $\sigma$ & $[N.m^{-1}]$ & $0.059$ & - \\ 
		\hline
		\end{tabular}
		\caption{Properties for saturated water at atmospheric conditions}
		\label{table:atm_water}
\end{table}

The initial temperature field was $T_{sat}$ in the liquid and a linear profile from the interface to the wall temperature. Three grid spacings were used on the $10 \, mm$ domain: $N_x=64,128,256$ with respective time step sizes of $t=0.002\,s$, $t=0.001\,s$ and $t=0.0005\,s$. The initial vapour layer thickness was taken as $322.5 \, \mu m$, which corresponds to a time $t=0.282435\,s$, if the vapour layer would have grown from zero thickness at time $t=0\,s$. The simulated time was $10s$.

\begin{figure}[!ht]
\begin{tabular}[width=0.95\textwidth]{cc}
\includegraphics[width=.47\textwidth]{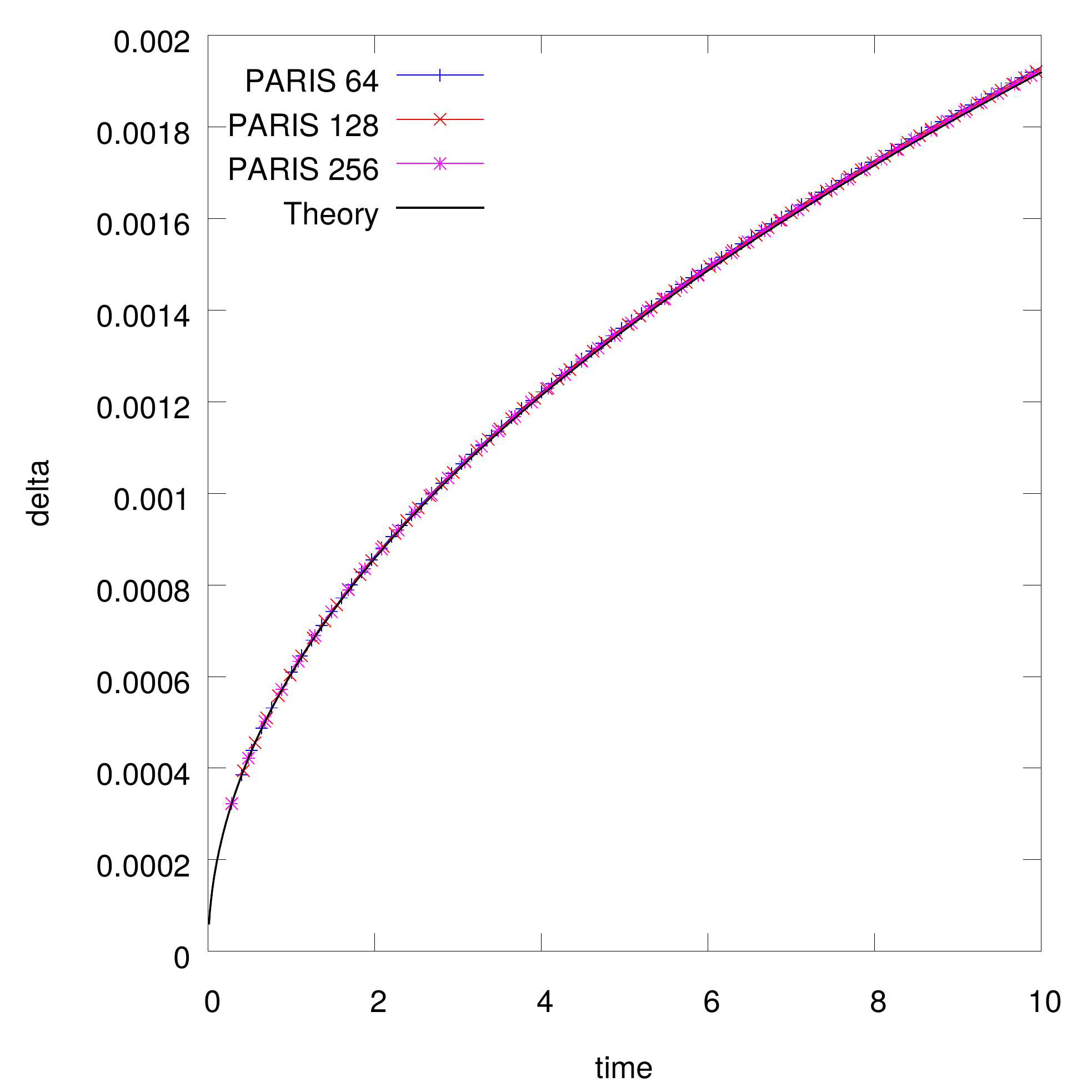} & 
\includegraphics[width=.47\textwidth]{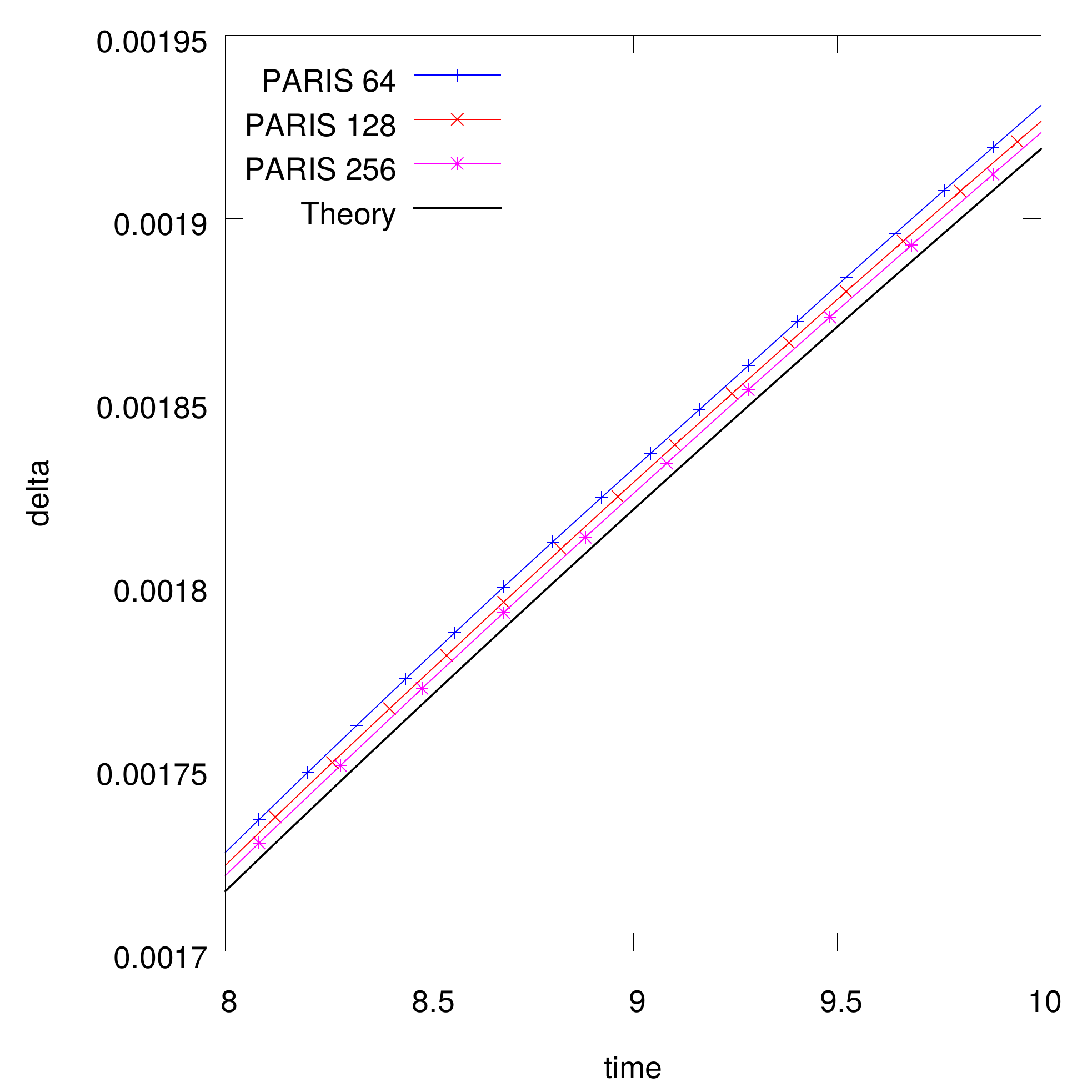}
\end{tabular}
\caption{The interface time evolution results for the Stefan problem.}
\label{fig:stefan_plot}
\end{figure}

The theoretical solution of the evolution of the interface thickness was obtained by solving \eqref{eq:stefan_lamda} numerically with a relative convergence error of $10^{-6}$ for $\lambda$. The results from \paris for the three grid resolutions are plotted in Fig.~\ref{fig:stefan_plot}. An excellent agreement was obtained. The relative errors of the interface position at $t=10\,s$ for the three grid resolutions are presented in Table \ref{table:stefan_errors}. The theoretical solution of \eqref{eq:stefan_delta} for $t=10\,s$ is $\delta(10)=1.919 \times 10^{-3} \,m$.

\begin{table}
	\centering
	\begin{tabular}[width=0.75\textwidth]{lcc}
		\hline
		\textbf{Grid points} & $\delta$ \paris & Relative error \% \\ 
		\hline
		$N_x=64$ & $1.93088 \times 10^{-3}$ & $0.62$ \\ 
		\hline
		$N_x=128$ & $1.92650 \times 10^{-3}$ & $0.39$ \\ 
		\hline
		$N_x=256$ & $1.92349 \times 10^{-3}$ & $0.23$ \\ 
		\hline
	\end{tabular}
	\caption{Relative errors for Stefan problem interface location at $t=10\,s$.}
	\label{table:stefan_errors}
\end{table}

The results show an error of less that $1$\% for the coarsest grid. Since the errors obtained here are small, the spatial convergence rate will be determined with a more demanding test case. 
 
\subsection{Bubble in superheated liquid}

A more demanding test case is that of a three-dimensional gas bubble at saturated temperature in a superheated liquid. The superheated temperature is denoted $T_\infty$ and no gravity force is applied. There is an analytical solution to the problem assuming spherical symmetry, obtained by Scriven \cite{Scriven59}. The solution gives the bubble radius $R$ as

\begin{table}[ht]
	\centering
		\begin{tabular}[width=0.75\textwidth]{lccc}
		\hline
		\textbf{Property} & \textbf{Units} & \textbf{Liquid}  & \textbf{Vapour} \\ 
		\hline
		Density $\rho$ & $\left[kg.m^3\right]$ & $2.5$ & $0.25$ \\ 
		\hline
		Specific heat $c_p$ & $\left[J.kg^{-1}.K^{-1}\right] $ & $2.5$ & $1.0$\\ 
		\hline
		Viscosity $\mu$ & $[Pa.s]$ & $7.0 \times 10^{-3}$ & $7.0 \times 10^{-4}$ \\ 
		\hline
		Thermal conductivity $k$ & $[W.m^{-1}.K^{-1}]$ & $0.07$ & $0.007$ \\ 
		\hline
		Surface tension $\sigma$ & $[N.m^{-1}]$ & $0.001$ & $0.001$ \\ 
		\hline
		Latent heat $h_{fg}$ & $\left[J.kg^{-1}\right] $ & $100.0$ & $100.0$ \\ 
		\hline
		\end{tabular}
		\caption{Properties for the bubble in superheated liquid case.}
		\label{table:supbubble_params}
\end{table}

\begin{equation}
R = 2 \, \beta_g \sqrt{\dfrac{k_\ell}{c_{p,\ell} \rho_\ell} t}
\end{equation} \,
with $t$ the time and the fluid properties as defined before. The value of $\beta_g$, sometimes referred to as the ``growth constant'' is obtained by solving

\begin{align}
&\dfrac{\rho_{\ell} c_{p,\ell} \left(T_\infty - T_{sat}\right)}{\rho_g \left(h_{fg}+\left(c_{p,\ell}-c_{p,g}\right) \left(T_\infty - T_{sat}\right)\right)} = \notag \\ 
&2 \, \beta_g^2 \int_0^1 exp \left( -\beta_g^2 \left( \left(1-\zeta\right)^{-2} - 2 \left(1 - \dfrac{\rho_g}{\rho_\ell} \right) \zeta -1 \right) \right) \, d\zeta
\end{align}

The solution is obtained numerically using the native numerical tools in the GNU package \code{Octave}. The temperature field is then given by

\begin{align}
T(r<R) &= T_{sat} \notag \\
T(r>R) &= T_\infty - 2 \, \beta_g^2 \left( \dfrac{\rho_g \left(h_{fg}+\left(c_{p,\ell}-c_{p,g}\right) \left(T_\infty - T_{sat}\right)\right)}{\rho_{\ell} c_{p,\ell}} \right) \notag \\
&\int_{1-\rfrac{R}{r}}^1 exp \left( -\beta_g^2 \left( \left(1-\zeta\right)^{-2} - 2 \left(1 - \dfrac{\rho_g}{\rho_\ell} \right) \zeta -1 \right) \right) \, d\zeta
\end{align}

The problem can be characterized by the Jacob number, given by

\begin{equation}
Ja = \dfrac{\rho_\ell c_{p,\ell} \left(T_\infty - T_{sat}\right)}{\rho_g h_{fg}} \,.
\end{equation}

A test case is defined with fluid properties given in Table \ref{table:supbubble_params}. Note that these properties do not correspond to a specific fluid, but was chosen for the purposes of a test with $\rfrac{\rho_\ell}{\rho_g} = \rfrac{\mu_\ell}{\mu_g} =  \rfrac{k_\ell}{k_g} = 10$. The heat capacity ratio is $\rfrac{c_{p,\ell}}{c_{p,g}}=2.5$.

The saturated temperature is $T_{sat}=1\,K$ with the temperature at infinity in the liquid $T_\infty=3\,K$. This results in $Ja=0.5$. A bubble of initial radius $R(t_0)=0.12$ is placed in the center of a unit cube. The initial temperature is taken from the analytical solution and the simulation is run for a time $t_f \approx 4 \times t_0$, with $t_f$ the final time.

\begin{table}[ht]
	\centering
	\begin{tabular}[width=0.75\textwidth]{lcc}
		\hline
		\textbf{Grid points} & Radius $R$ \paris & Relative error \% \\ 
		\hline
		$N=64^3$ & $2.279 \times 10^{-1}$ & $5.56$ \\ 
		\hline
		$N=128^3$ & $2.361 \times 10^{-1}$ & $2.15$ \\ 
		\hline
		$N=256^3$ & $2.404 \times 10^{-1}$ & $0.37$ \\ 
		\hline
	\end{tabular}
	\caption{Relative errors for the bubble radius $R$ at $t=4 \, t_0$.}
	\label{table:superbub_errors}
\end{table}

\begin{figure}[ht]
\centering
\includegraphics[width=.55\textwidth]{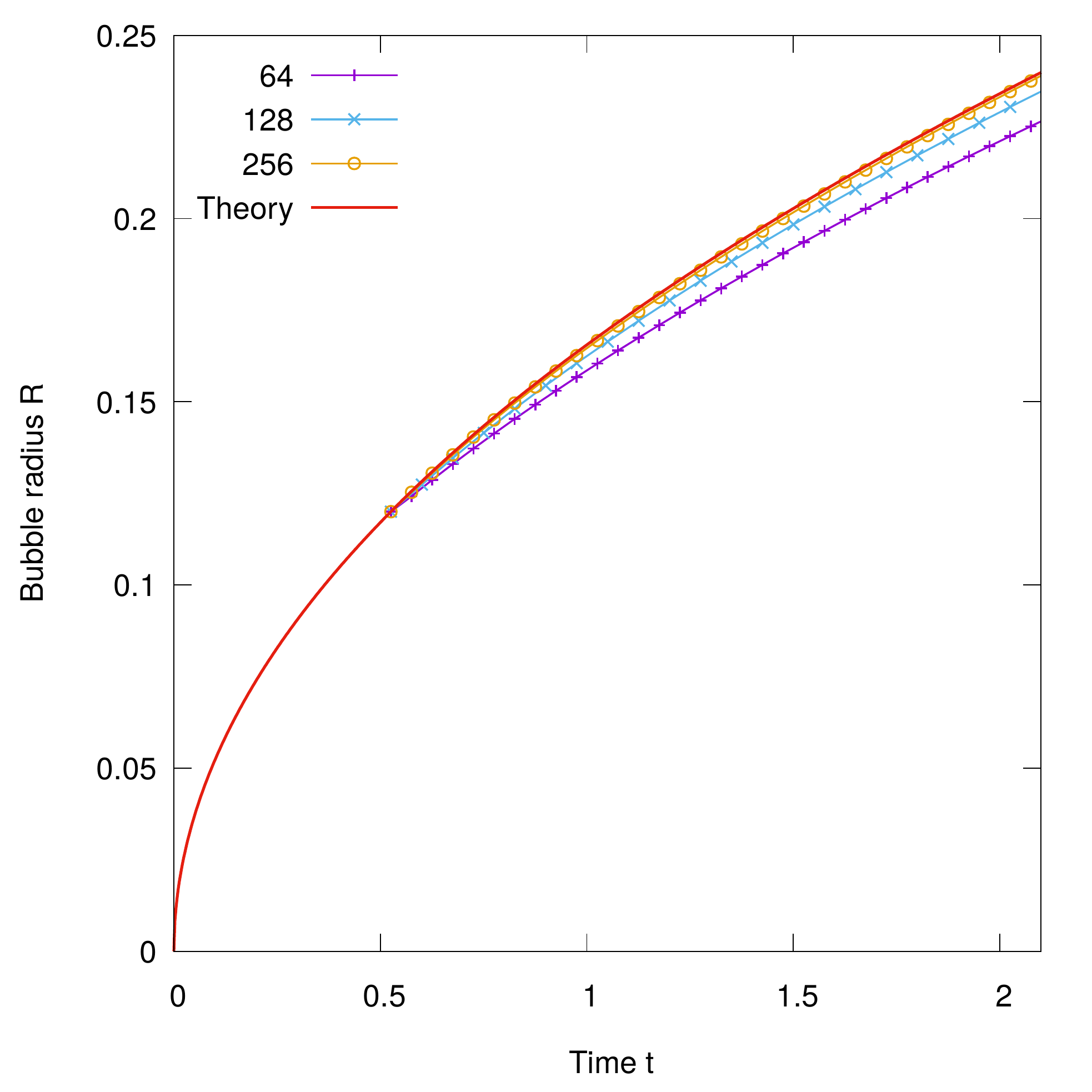}  
\caption{The time evolution of the bubble radius for a bubble at saturated temperature in a superheated liquid.}
\label{fig:SupBubble}
\end{figure}

Three test cases were run: $N=64^3,128^3,256^3$ with respective time step sizes of $t=0.01\,s$, $t=0.005\,s$ and $t=0.0025\,s$. The time evolution of the bubble radius in the simulations is compared to the theoretical value in Fig.~\ref{fig:SupBubble}. The theoretical value of the bubble radius at the final time was determined by the analytical solution: $R(t_f=2.2156\,s)=0.2413$ and compared to simulation. The results are shown in Table \ref{table:superbub_errors}, along with the relative error compared to the theoretical value. The largest error was 5.56\% on the coarsest grid.

\begin{figure}[ht]
\centering
\begin{tabular}[width=1.00\textwidth]{ccc}
\includegraphics[width=.30\textwidth]{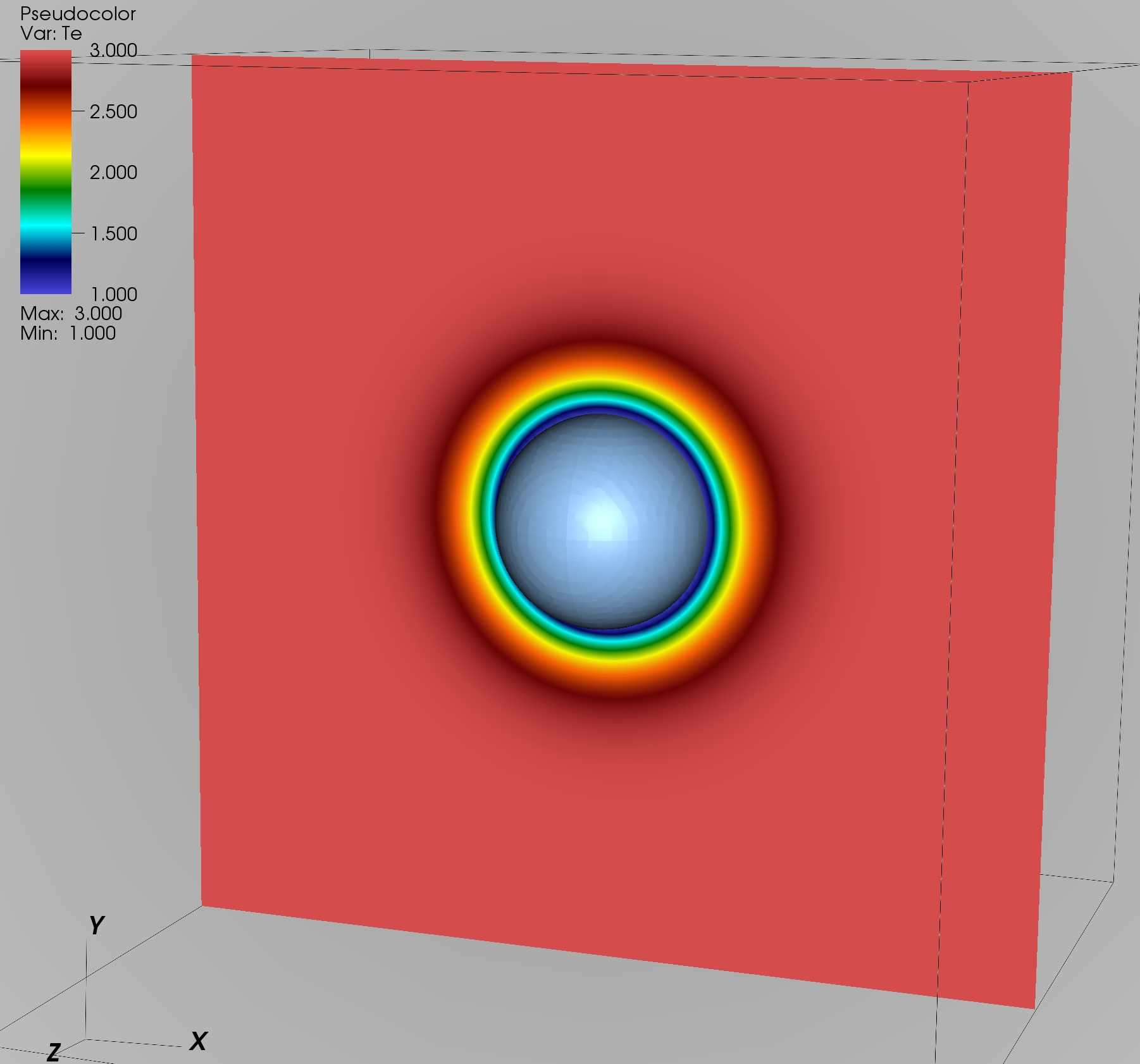} & 
\includegraphics[width=.30\textwidth]{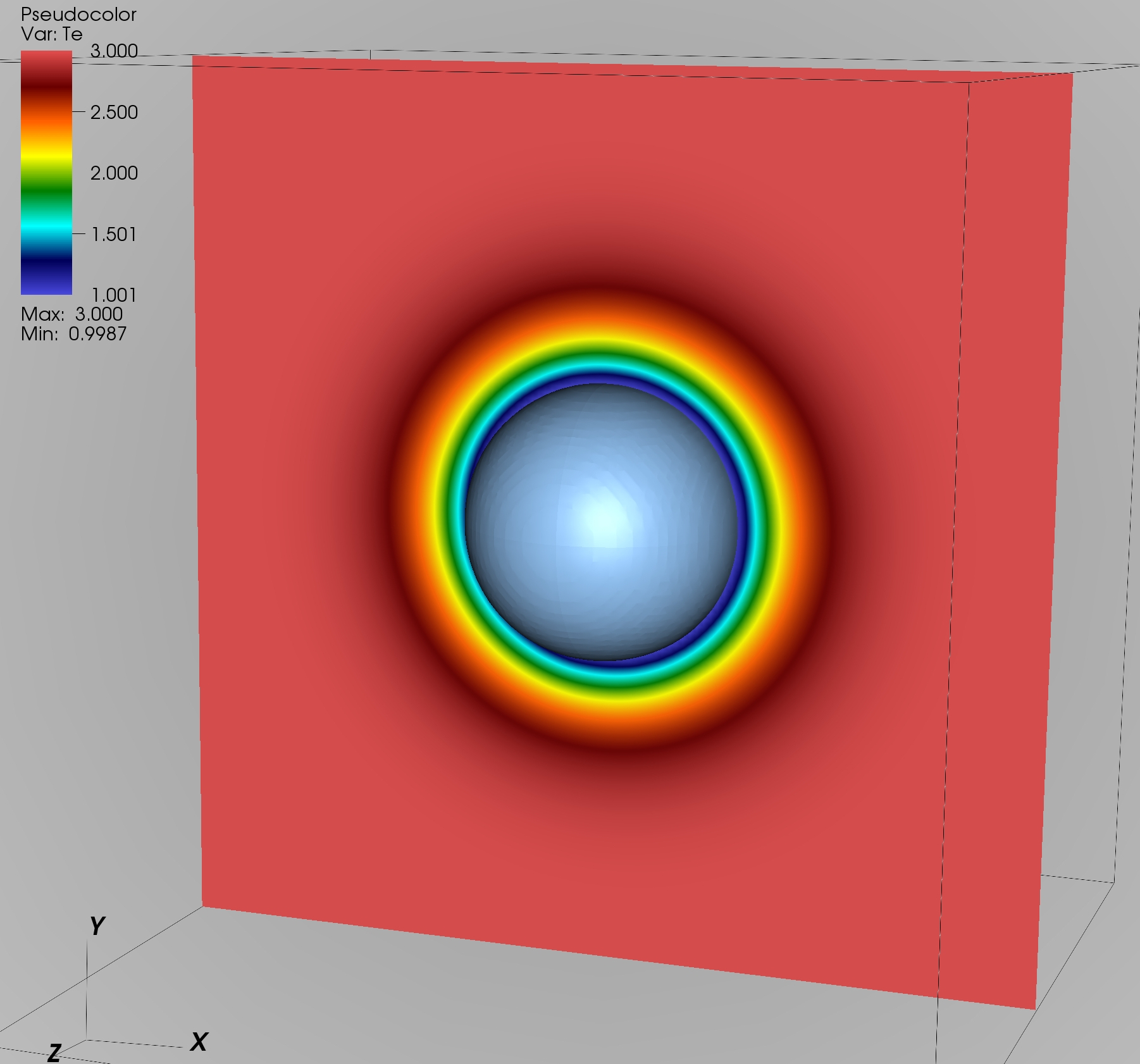} &
\includegraphics[width=.30\textwidth]{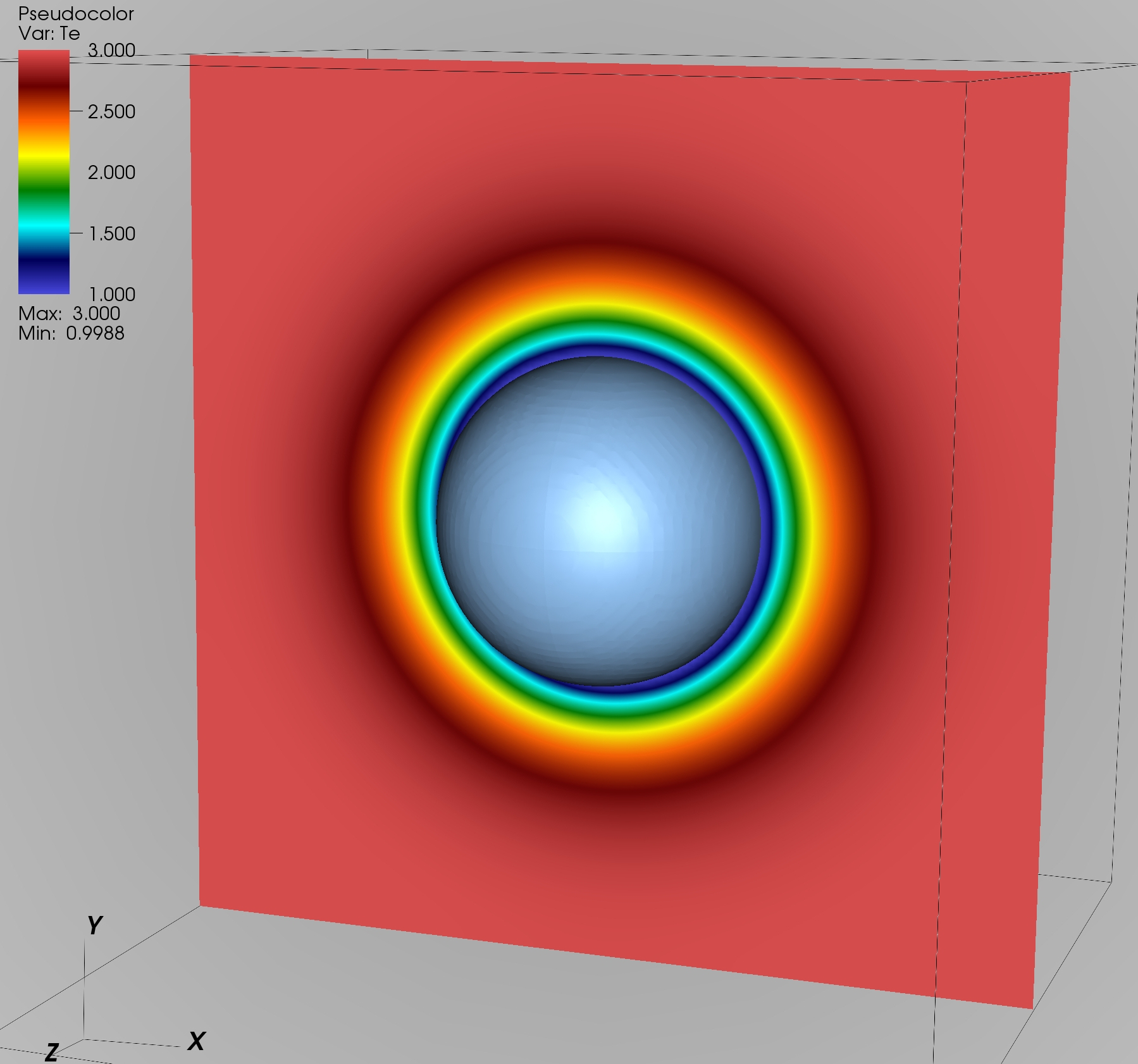} \\
\includegraphics[width=.30\textwidth]{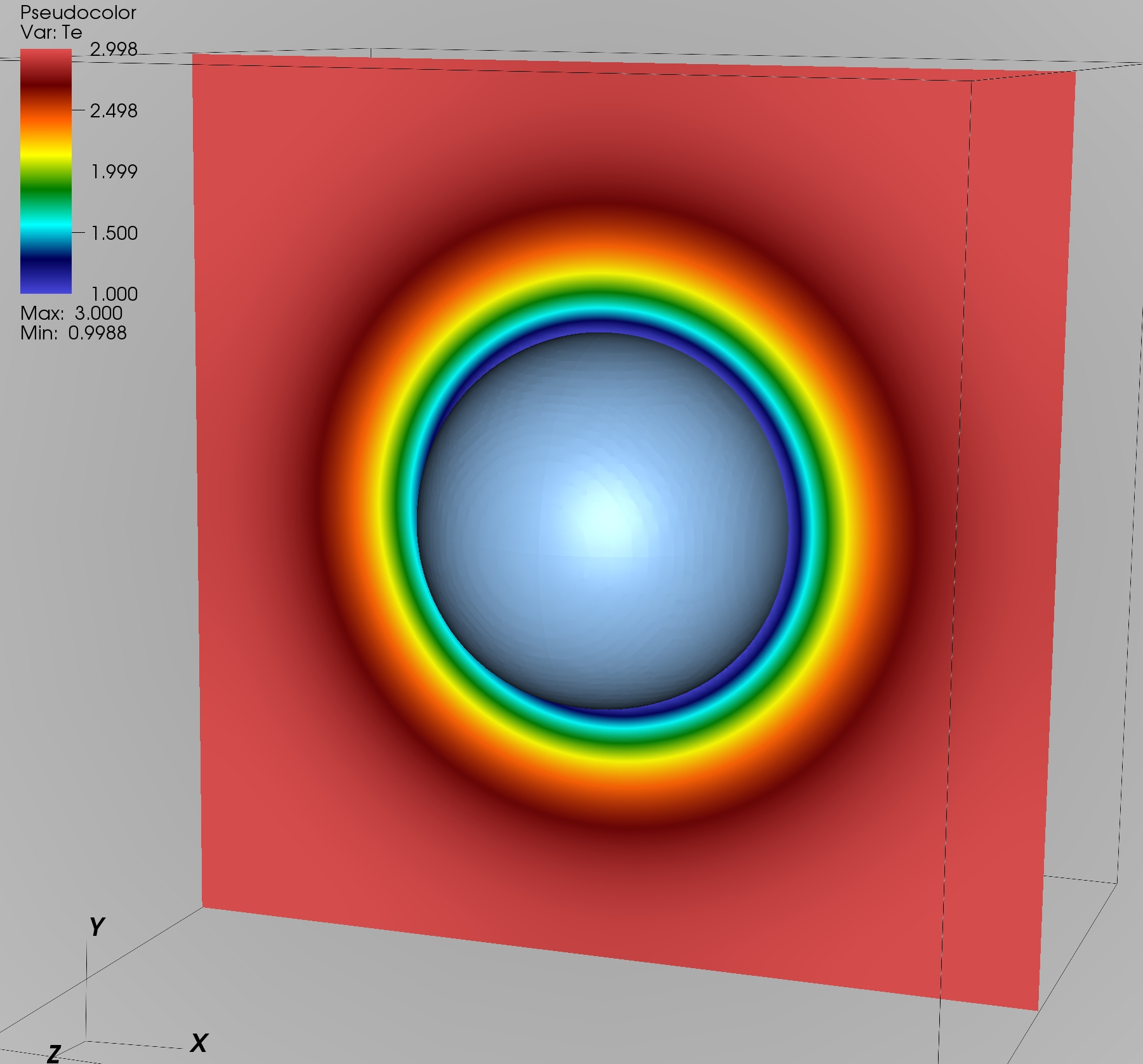} &
\includegraphics[width=.30\textwidth]{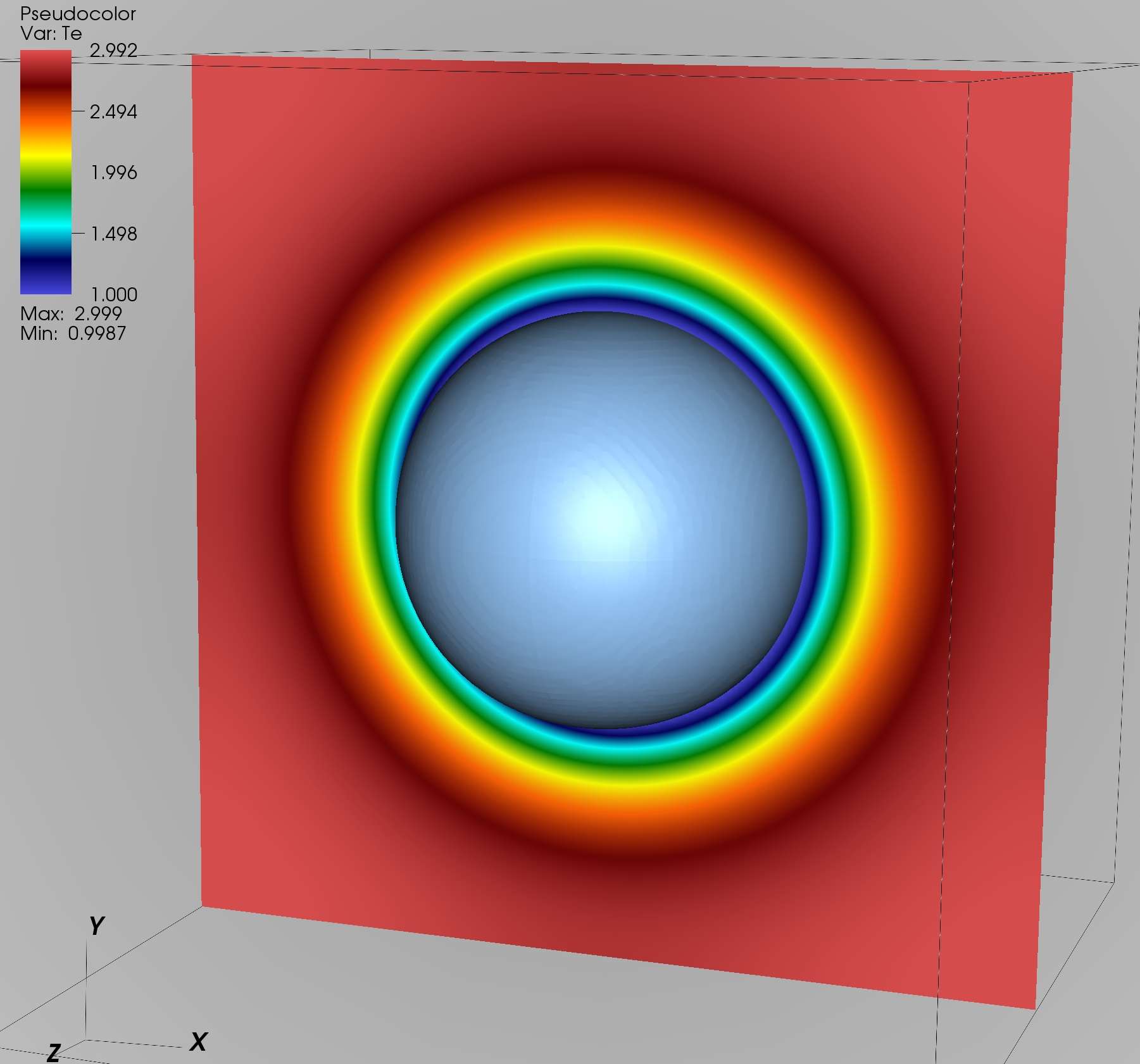} &
\includegraphics[width=.30\textwidth]{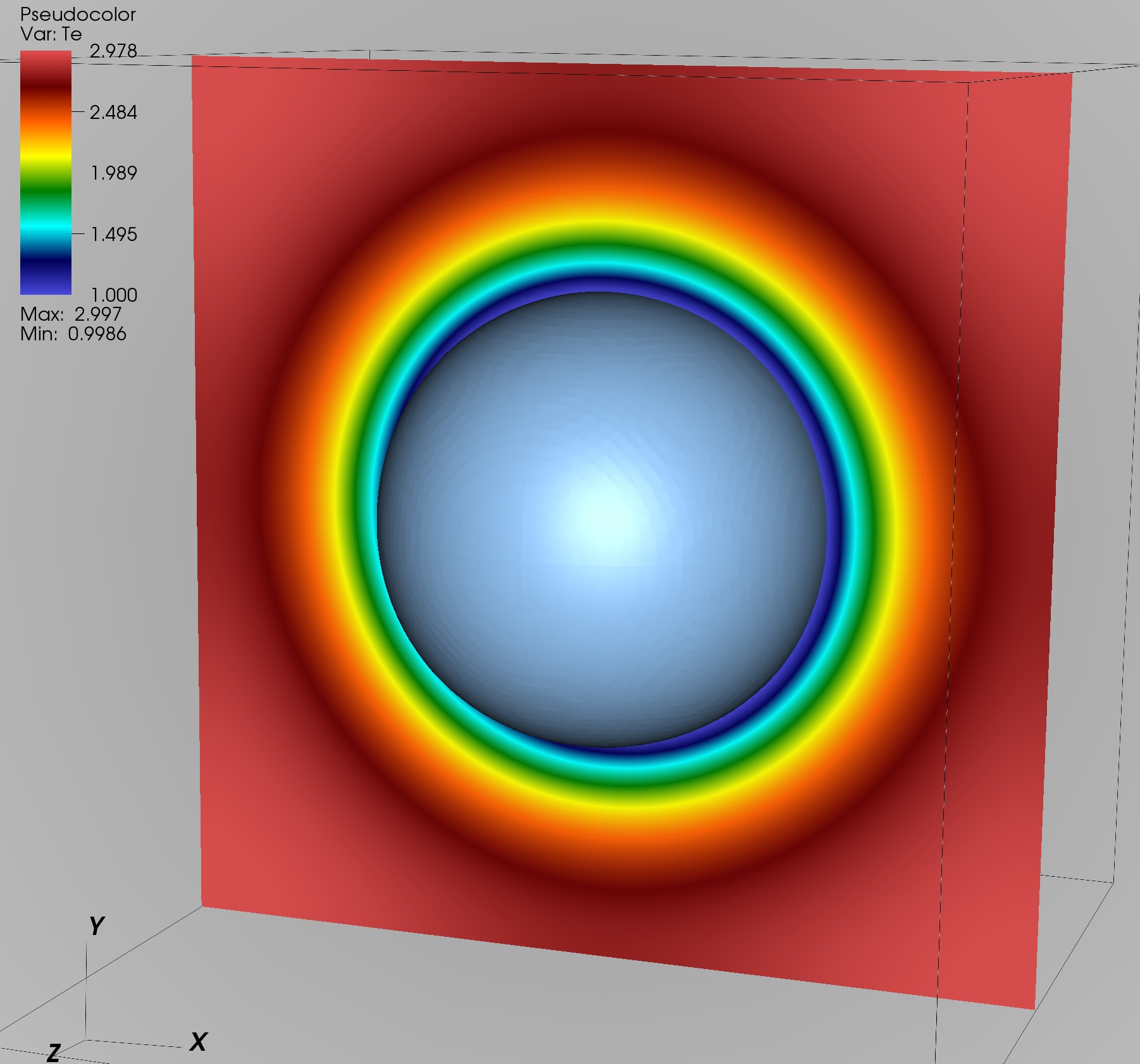}
\end{tabular}
\caption{A time sequence presentation of a vapor bubble in superheated liquid, taken at $0.4$s intervals starting at $t=0$s at the top left. The blue surface indicates the $c=0.5$ iso-surface, which is a representation of the interface. The colour scale shows temperature at a mid-plane through the bubble.}
\label{fig:SupBub_seq}
\end{figure}

One important source of the error is the resolution of the thermal boundary layer in the superheated liquid. Lower resolution results in an inaccurate calculation of the temperature gradient in the liquid, which leads to an inaccurate mass transfer rate, \eqref{eq:mdot_calc}. A time sequence of the simulation is shown in Fig.~\ref{fig:SupBub_seq}, indicating the bubble growth and superheated temperature around the interface. It is encouraging to notice that the thermal boundary layer as well as the bubble shape remains spherically symmetric along with a relatively accurate prediction of the bubble radius evolution.

\section{Conclusion}

The Direct Numerical Simulation of flows with phase change was studied via a geometric PLIC VOF scheme. The discontinuity in velocity across the interface due to phase change poses a particular challenge to conventional VOF schemes that rely on a smooth, divergence--free velocity field. In this work a novel two-step VOF advection method was proposed. In the first step, the interface is advected with a divergence free liquid velocity, which is obtained from the solution of a Poisson problem that decomposes the one-fluid velocity into a liquid velocity and a phase-change component. In the second step, the phase change component is accounted for with an explicit interface shift in the local normal direction.

The geometric treatment of VOF advection is applied consistently to the thermal energy advection term. The thermal energy diffusion term is solved implicitly, using a similar technique to Sato and Ni\v{c}eno \cite{Sato13}, where an asymmetric stencil is used to apply the interface temperature directly.

The method was implemented in \paris (with full parallel computation capability) and was tested using three benchmark test cases. The two-step VOF advection method was tested on a two-dimensional evaporating droplet. The results on three different grids were indistinguishable and showed excellent agreement to the $\rfrac{1}{t^2}$ evolution of the volume. The one-dimensional Stefan problem was solved to excellent accuracy. A three dimensional bubble at saturated temperature in a superheated liquid with $Ja=0.5$ was simulated at three different grid resolutions. It was found that the spherical shape was preserved and the radius evolution was predicted with good accuracy, especially in the high resolution case where the temperature gradient in the thermal boundary layer could be captured with greater accuracy than in the cases with lower resolution.

\section*{Acknowledgements}

This work is based on research supported by the National Research Foundation of South Africa (Grant Numbers: 89916). The opinions, findings and conclusions or recommendations expressed is that of the authors alone, and the NRF accepts no liability whatsoever in this regard. The authors thank Roy Horwitz for the creation of a numerical solver for the theoretical bubble radius and temperature profile in the flashing saturated bubble test case that was used to validate the numerical results.

Screen shots presented in this work are from \textsc{VisIt} \cite{visit}.

\bibliography{ms.bbl}
\bibliographystyle{siam}

\end{document}